%% file: article_main.tex
\documentclass[fleqn,usenatbib]{mnras}

\usepackage{newtxtext,newtxmath}

\usepackage[T1]{fontenc}

\DeclareRobustCommand{\VAN}[3]{#2}
\let\VANthebibliography\thebibliography
\def\thebibliography{\DeclareRobustCommand{\VAN}[3]{##3}\VANthebibliography}

\usepackage{chemformula}
\usepackage{siunitx}
\usepackage{subcaption}

\usepackage{graphicx}	
\usepackage{amsmath}
\usepackage[version=4]{mhchem}

\title[Nitrogen chemistry of hycean worlds]{Nitrogen chemistry of hycean worlds on the example of K2-18b}

\author[M. W. Radecka et al.]{
Maja W. Radecka,$^{1}$\thanks{Current address: Multiphase Chemistry Department, Max Planck Institute for Chemistry, Hahn-Meitner Weg 1, 55128 Mainz, Germany. E-mail: mr865@cantab.ac.uk, m.radecka@mpic.de}
Paul B. Rimmer,$^{1}$\thanks{E-mail: pbr27@cam.ac.uk}
\\
$^{1}$Cavendish Laboratory, University of Cambridge, Cambridge CB3 0HE, UK\\ 
}

\date{Accepted XXX. Received YYY; in original form ZZZ}

\pubyear{\the\year{}}

\begin{document}
\label{firstpage}
\pagerange{\pageref{firstpage}--\pageref{lastpage}}
\maketitle

\begin{abstract}
A recent observation of the exoplanet K2-18b sparked interest among scientists - large amounts of carbon dioxide and methane were detected in an \ch{H2}-rich background atmosphere. If the planet is a hycean world (liquid water ocean + hydrogen-dominated atmosphere), it could be habitable under certain conditions. The presence of carbon, hydrogen and oxygen was already confirmed, however, there was no detection of nitrogen or its compounds. Molecular nitrogen is difficult to detect directly. This study concentrates on possible photochemical products of \ch{N2} such as \ch{HCN}, \ch{NH3} and \ch{HC3N}. We set approximate limits on the amount of nitrogen bearing species by varying atmospheric parameters, such as the Eddy Diffusion coefficient and the amount of \ch{N2} present from 10 ppm to 10\%. If the bulk nitrogen-containing gas in the atmosphere is \ch{N2}, photochemistry produces only trace amounts of the aforementioned species. However, if ammonia is the main source of nitrogen, then the quantities of \ch{NH3}, \ch{CH5N} and HCN approach detectable range. \ch{HC3N} and NO are bad tracers of the nitrogen source in the atmosphere, because they are produced in similar amounts in all tested scenarios. Assuming equilibrium chemistry at the surface of K2-18b results in underprediction of \ch{CO2} abundance. This result combined with the non-detection of ammonia by JWST suggests the planet is not a typical sub-Neptune, but could be indeed a hycean world or magma ocean planet. We also found that \ch{C2H6} is produced in significant amounts - if it is detected in the future, it could serve as a proxy for DMS presence. 
\end{abstract}




\input{Introduction}
\input{Methods}
\input{Results}
\input{Discussion}
\input{Conclusions}
\section*{Acknowledgements}
\input{Acknowledgments}

\section*{Data Availability}
The data is available upon request to the corresponding author. 



\bibliographystyle{mnras}
\bibliography{references} 



\appendix
\onecolumn
\input{Appendix_1}
\input{Appendix_5}
\input{Appendix_4}


\bsp	
\label{lastpage}
\end{document}

%% file: Introduction.tex
\section{Introduction} \label{sec:intro}

Studying the properties of an exoplanet by observing its transit is an effective way to get an insight into the composition of its atmosphere (\cite{Sara_textbook}). Large transit depth can be obtained by observing bigger planets orbiting smaller stars (\cite{Madhusudhan_2018}). Additionally, if the atmosphere of such planet has low mean molecular weight, retrieved transmission spectrum is stronger. These qualities draw attention towards sub-Neptunes orbiting M-dwarf stars. Recent surveys, detected large numbers of such planets, in which size category falls K2-18b (\cite{Montet_2015}, \cite{Cloutier_2017}) - a planet of mass $M_p = 8.63 M_{\earth}$ and radius $R_p = 2.61 R_{\earth}$ (\cite{Cloutier_2019}). 

An important question to ask is whether or not the planet could be habitable. The planet needs to lie within the habitable zone of its star and have significant abundance of organic material. The environment on the planet can be drastically different from Earth, but it makes sense to look for familiar byproducts of biological activity and building blocks for simple cellular organisms (like the precursors of amino acids). Organic chemistry that we are familiar with is almost solely based on carbon, hydrogen, nitrogen and oxygen. These molecules are also present in Earth's atmosphere as oxygenated gases or are trapped in aerosols, hence it would be expected to find them in exoplanetary atmospheres as well. 

Liquid water is believed to be necessary for life to form. \cite{Madhusudhan_2021} show how liquid oceans could be sustained on hycean planets, even outside of the conventional terrestrial habitable zone. However, the planet does not need to have vast oceans to sustain life - organic molecules could form on dry land or be delivered by impacts and later shielded from harmful radiation by the planet's magnetic field. Since for now, we only know one habitable planet - the Earth - we notice that nitrogen in the atmosphere seems possibly important for habitability. It is good to get an idea about its abundance on K2-18b.

What we can detect and how we can detect it, is another question that needs to be answered. Molecular nitrogen does not have a dipole moment, therefore it is hard to observe directly through transmission spectroscopy. To find alternative ways of estimating its total abundance on the planet, we could look for nitrogen bearing molecules, such as ammonia or other photochemical products of nitrogen. This study will focus on determining the relationship between the total amount of nitrogen in the atmosphere, its source gas, and its compounds. 

An additional issue to resolve is to determine the class of the observed planet. Due to different chemical and physical conditions, the presence or detection of some species can be impossible, depending on the planetary type. Upon the discovery of K2-18b, researchers have been arguing over its planetary type - the easiest association would be to classify it as a mini-Neptune, however \cite{Madhusudhan_2020} argued that the planet could be a hycean world - a new category of an exoplanet containing liquid ocean and hydrogen dominated atmosphere. A model of K2-18b with an ocean and hydrogen envelope was also explored in \cite{Hu_2021}.
Due to the non-detection of ammonia in its atmosphere, \cite{Madhusudhan_2023} claim K2-18b is less likely to be a rocky world with a thick hydrogen envelope or a mini-Neptune. In Section \ref{intro1} we provide a brief summary of studies on K2-18b and in Section \ref{intro2} we discuss the importance of nitrogen for life to develop on a planet.

\subsection{K2-18b as a hycean world?}\label{intro1}

Hycean worlds posses a couple of desirable qualities (\cite{Chemical_Hycean}, \cite{Madhusudhan_2021}): 
    (1) hycean habitable zone spans much wider than the terrestrial habitable zone, hence there are more possibilities for life to develop;
    (2) hycean planet would have larger atmospheric scale height due to low molecular weight of hydrogen and hence it would be more favorable for transmission spectroscopy.

Before we treat K2-18b as a hycean world, we need to point out a few challenges associated with this classification. \cite{Innes_2023} showed how the hydrogen collision induced absorption: (\ch{H2}-\ch{H2} CIA - \cite{BORYSOW1992169}) would affect the planetary heat balance, increasing the planet's mean temperature and pushing habitable zone further from the star. On collision between hydrogen molecules, a temporary dipole is induced, that can absorb IR radiation and increase opacity of the atmosphere. CIA will be more effective deeper in the atmosphere, where the density is higher and collisions of molecules are more frequent, causing surface temperature to rise. Similarly, \cite{Scheucher_2020} previously pointed out, that the presence of substantial liquid water reservoir would result in water vapor dominated atmosphere with increased mean molar mass and hence decreased scale height, no longer fitting the conditions characterizing the hycean planet. To counteract these effects, significant hazes should be present in the atmosphere to reflect back some of the radiation and allow the surface to be habitable (\cite{Madhusudhan_2023}). However, one needs to be careful when assuming a straightforward contribution of haze, as \cite{He2024} has shown that varying the optical properties of haze can modify the spectrum (even in the JWST range). 

Non-detection of water could mean that the stratosphere is dry (just like on Earth). According to \cite{Chemical_Hycean}, the depletion of ammonia relative to methane could hint a presence of a liquid ocean. However, if there were both ammonia and the liquid ocean, it could mean there is an additional flux of \ch{NH3}, possibly due to life or chemical equilibrium at the surface. The assumption of chemical equilibrium at the bottom of the atmosphere (BOA) is closely related to the pressure and temperature at the surface. As \cite{Yu_2021} shows: cool, shallow surface ($\mathrm{p_{BOA} < 10}$ bar) can interfere with thermochemical recycling which ends in the destruction of photochemically fragile species (\ch{NH3}, \ch{HCN}, \ch{CH4}, \ch{C2H2}, \ch{H2O}) and buildup of stable species (\ch{CO2}, \ch{CO}, \ch{N2}, \ch{H2} and \ch{H2O}). Since each specie approaches equilibrium at a different pressure, surface placement in our model will cause equilibrium for some and disequilibrium for other molecules. \cite{Hu_2021} finds that \ch{NH3} is depleted in small atmospheres, but their findings also point out that increasing the metallicity of the planet to 100x solar may give a very similar spectrum: strong \ch{CO2} and \ch{CO} features, but an absence of \ch{NH3} or \ch{HCN}. Similar results are obtained by \cite{Tsai_2021}, as they find that the presence of ammonia and methanol (\ch{CH3OH}) is clear representative of surfaces in sub-Neptunes.

Apart from hazes, clouds also affect the overall planet's albedo, impact the circulation, condensation, and the abundance of other species in the atmosphere. Their presence can block the upward transport of some molecules through the uptake by the cloud droplets and hence they will not be visible in the retrieved spectrum. \cite{Benneke_2019} reports that K2-18b atmospheric models with clouds in the mid-atmosphere are favoured by 2.6$\sigma$ and those without contain unrealistic amounts of water vapour (\cite{Scheucher_2020}, \cite{Innes_2023}). They also constrain the maximum abundances of \ch{N2} to 10.9\%, \ch{CO2} - 2.4\%, \ch{CO} - 7.45\%, \ch{NH3} - 13.5\% and \ch{CH4} - 0.248\%. All of these seem plausible, however the retrieved amount for methane in \cite{Madhusudhan_2023} is larger - 0.913\%. Nevertheless, \cite{Blain_2021} already predicted that in the spectrum, methane should dominate over water. 

\cite{Wogan_2024} tests 3 models for K2-18b: lifeless hycean, inhabited hycean and mini-Neptune with 100x solar metallicity. The first model is disregarded, because it does not fit the data well, however the last two present very similar results and goodness of fit. The paper uses analogous parameters to this research: UV flux of star GJ-176, model with 100x solar metallicity and a temperature profile with a moist pseudo adiabat in the deeper atmosphere and isothermal stratosphere. 
However, \cite{cooke2024considerationsphotochemicalmodelingpossible} argue that inhabited hycean case better fits the data, as does a more recent paper by \citet{hu2025waterrichinteriortemperatesubneptune}.
Additionally, \cite{Piette_2020} show that increasing metallicity increases planet's temperature - testing different metallicity values against the recovered spectra would help finding the optimal temperature profile for the planet. 
Finally, there is a real possibility that the planet is a magma ocean world, and dissolution properties of different volatiles could in principle explain the same chemical signatures we see \citep{Shorttle2024}.

\subsection{Nitrogen in the atmosphere}\label{intro2}

Nitrogen is one of the most important elements for the development of life on Earth. It is essential for formation of RNA, DNA, proteinogenic amino acids and their precursors. Creation of such precursors and their delivery to different planets is a topic of many ongoing studies. Nitrogen bearing compounds can simply be the products of abiotic chemistry in the atmosphere or could be considered exoplanetary biosignatures if they fulfill criteria listed below.

An ideal biosignature gas should not exist naturally in the atmosphere, needs to have a strong spectral feature and should not be produced by any geochemical process or photochemistry \citep{Sara_textbook}. Review by \cite{schwieterman2017a} describes which molecules are good biosignatures in what conditions and why certain molecules should not be considered as biosignatures, even though they can be produced by life. \cite{Hu_2013} shows how organic haze could imply a certain \ch{CH4} to \ch{CO2} ratio on the planet and similarly sulfur aerosols - \ch{S8}, \ch{H2SO4}, could provide a good gauge on the surface amounts of these molecules and hence serve as a proxy for a presence of biomass. \cite{Seager_2013b} develops a model to check if the detection of certain biomarkers will imply plausible amounts of biomass present and finds that \ch{CH4}, \ch{H2S} and \ch{N2O} will most likely be a false positive. Possible biosignature gases include ammonia in \ch{H2} rich atmosphere, because direct reaction \ch{N2 + 3 H2} $\rightarrow$ \ch{2 NH3} has an energetic barrier (\cite{Seager_2013a}, \cite{Huang2022}). \ch{DMS} (\ch{(CH3)2S}) or \ch{CH3Cl} are good proxies for life, however they are usually produced in trace amounts and the destruction of \ch{DMS} by UV radiation further reduces its abundance. Nevertheless, it could be inferred from the increased quantities of \ch{C2H6} relative to the amount of methane present - \cite{Domagal_2011} argues it could be produced by \ch{CH3} radicals coming from the \ch{DMS} dissociation. 

Molecules often implicated in prebiotic chemical scenarios are ammonia (\ch{NH3}), hydrogen cyanide (\ch{HCN}) and cyanoacetylene (\ch{HC3N}) (\cite{Claringbold_2023}, \cite{Rimmer_Ranjan}). 
The delivery options for them include cometary impacts, volcanism, lightning and photochemistry. \cite{Rimmer_2021} predicts that \ch{HC3N} should be observable in hydrogen rich atmospheres with high Eddy Diffusion coefficient ($10^6 < K_{zz} < 10^{10}$ \unit{cm^2 s^{-1}}) - similar results are also found in this paper. Cyanoacetylene abundance is interconnected with the presence of hydrogen cyanide in the atmosphere through reactions like: \ch{HCN + 2 CH4 } $\rightarrow$ \ch{ HC3N + 4 H2}. The C/O ratio in the exoplanet's atmosphere is closely related to the content of \ch{HCN}, which is found to be produced more easily in methane or acetylene rich atmospheres \citep{Rimmer_2019}. 

Lightning causes stable and chemically inert molecules of \ch{N2} to break apart and release N atoms and ions, which can then react with other constituents of the atmosphere. Similar effect can be obtained by stellar flares (\cite{Airapetian_2016}). \cite{Rimmer_2016} discusses how in a reducing atmosphere lightning could bring enough help to overcome the energetic barriers and produce glycine (\ch{NH2CH2COOH}), while \cite{Ioppolo_2021} presents a formation pathway for glycine on comets in the interstellar medium, which could be later delivered by an impact. There are even more ways of producing basic life building blocks in the atmosphere or the interstellar medium. \cite{Stark_Helling_Diver_Rimmer_2014} show how electrostatic forces on dust grains can initiate formation of certain prebiotic molecules. Or a combination of \ch{SO3^{2-} + HCN} could lead to RNA precursors (\cite{Rimmer_2018}). 

There are many ways in which nitrogen could be present or delivered to the atmosphere. Among possible sources of atmospheric nitrogen are volcanism, lightning, photochemistry or nitrogen sequestration in the ocean. Nitrogen contributes to life development and habitability of the planet through being a greenhouse gas (like \ch{N2O}) and building amines and amino acids. Most abundant nitrogen-bearing molecule will depend a lot on the surface conditions: what is the pressure? is there an ocean? are there nitrogen fixating bacteria present? 

The focus of this paper is to establish how the concentration of detectable nitrogen-bearing compounds can be used to distinguish whether the bulk nitrogen-containing gas in the atmosphere is \ch{N2} or \ch{NH3}.

\subsection{Paper structure}

In Section \ref{sec:methods}, we outline the functionalities of our model and the principles behind our choice of input variables. In Section \ref{sec:results} we will present the results of our simulations and in Section \ref{sec:discussion} we will discuss these results. The overall conclusion of this research can be found in Section \ref{sec:conclusions}.

%% file: Methods.tex
\section{Methods}\label{sec:methods}
This section contains illustrations of methods applied in this study. It begins with a short description of the model used (section \ref{subsec:ARGO}) and follows with justification of choices of the input parameters in section \ref{subsec:inputs}, such as mixing ratios of certain species or the stellar spectrum used (section \ref{subsec:opt-depth}). At the end, thermodynamic parameters of the atmosphere are established, such as the pressure-temperature profile (section \ref{subsec:pT-profile}) and the speed of Eddy turbulent mixing (section \ref{subsec:Kzz}). 

\subsection{ARGO: 1D Photochemistry \& Diffusion Code}\label{subsec:ARGO}
This study uses ARGO - the code developed by \cite{Rimmer_2016}. ARGO is a photochemistry 1D diffusion model that follows Lagrangian formulation of molecular transport. It uses a 1D continuity equation for a chemical species $i$ in vertical, time-dependent transport:
\begin{equation}
    \frac{\partial n_i}{\partial t} = P_i - L_i - \frac{\partial \Phi_i}{\partial z}
\end{equation}
where $n_i [\unit{cm^{-3}}]$ is the number density of species $i$, $P_i[\unit{cm^{-3} s^{-1}}]$ is the rate or production and $L_i[\unit{cm^{-3} s^{-1}}]$ is the rate of loss of species $i$. $\Phi_i[\unit{cm^{-2} s^{-1}}]$ is the flux (molecular and Eddy Diffusion combined).

The model tracks a parcel of gas moving through the atmosphere, following a fixed pressure-temperature profile. At each height/pressure level, it records the mixing ratios of all molecules present. Once the parcel reaches the top of the atmosphere, it follows the same path to the bottom of the atmosphere, again recalculating all of the mixing ratios and reaction rates. Radiative transfer model is applied between each global run (\cite{bangera2025kineticphotochemicaldisequilibriumpotentially}). This process is repeated until model converges to an equilibrium. For this case, it usually took between 6-9 global runs to reach the desired accuracy.

This study is using the updated version of Chemical Network - Stand-2023May (\cite{rimmer2021hydroxidesaltscloudsvenus}, \cite{Hobbs_2021} - see supplementary material for the full network). It consists of 6623 reactions, involving stable species, ions, as well as condensing and vaporizing species. For our project, we have only focused on reactions involving molecules with H, C, N, O and S. For more details about the model development and the chemical network, please see \cite{Rimmer_2016}.

\subsection{Choosing input parameters}\label{subsec:inputs}

Model inputs include surface abundance of species, pressure-temperature profile and stellar flux. The reference inputs contain the JWST data from \cite{Madhusudhan_2023} (Table \ref{tab:reference_sample}). The presence of ocean at the surface implies there is water vapour in the lower parts of the atmosphere. However, \ch{H2O} was not detected by JWST, which implies a dry stratosphere. In order to achieve that, we have simulated K2-18b with a dry atmosphere to replicate these conditions. There is an additional benefit to modeling a dry atmosphere - in the samples with water present, nitrogen species were depleted faster, hence the results we present here could be considered "the best case scenario for nitrogen". The comparison between dry and wet samples is shown in the Appendix \ref{Appendix_5}. Other elements, such as nitrogen and sulfur, were added in quantities ranging from trace amounts to being one of the dominant species in the atmosphere (see Table \ref{tab:reference+nitrogen}). A case with 100x solar metallicity was also considered and it was commonly tested in literature (\cite{Wogan_2024}, \cite{Blain_2021}, \cite{Soni_2023}). In this sample, ammonia is the dominating source of nitrogen in the atmosphere (Table \ref{tab:100x_solar}). The choices of given parameters are justified in the sections below: p-T profile (section \ref{subsec:pT-profile}), Eddy Diffusion coefficient (section \ref{subsec:Kzz} and stellar flux (section \ref{subsec:opt-depth}).

\begin{table}
    \centering
    \begin{tabular}{||ccc||}
    \hline
    \multicolumn{3}{||c||}{\textbf{Reference sample}} \\
    \hline\hline
       \textbf{ \ch{H2}} & \textbf{\ch{CO2}} &   \textbf{\ch{CH4}}\\
        \hline
          97.325\% & 1.768\% & 0.907\% \\
    \hline
    \end{tabular}
    \caption{\textbf{Reference sample} - these inputs do not change throughout all of \ch{N2} sample measurements (see subsection \ref{subsec:n2} and Table \ref{tab:reference+nitrogen}). Values for \ch{CO2} and \ch{CH4} are taken from Table 2 (No Offset values) in \protect\cite{Madhusudhan_2023} and adjusted, so that the rest of the atmosphere is filled with molecular hydrogen.}
    \label{tab:reference_sample}
\end{table}

\begin{table}
    \centering
    \begin{tabular}{||ccc||cc||}
    \hline
    \multicolumn{3}{||c||}{\textbf{Reference sample}} & & \textbf{SAMPLE} \\
    \hline\hline
       \textbf{\ch{H2}} & \textbf{\ch{CO2}} &  \textbf{\ch{CH4}} & \textbf{\ch{N2}} & \\
         \hline\hline
         88.533\% & 0.825\% & 1.608\% & 9.034\% & A \\
          92.721\% & 0.864\% & 1.684\% & 4.731\% & B \\
          95.430\% & 0.889\% & 1.733\% & 1.948\% & C  \\
          97.368\% & 0.898\% & 1.750\% & 0.983\% & D  \\
          97.229\% & 0.906\% & 1.766\% & 0.099\% & E  \\ 
          97.315\% & 0.907\% & 1.768\% & $9.93\times10^{-5}$ & F \\
          97.325\% & 0.907\% & 1.768\% & $9.93\times10^{-6}$ & G\\
         \hline
    \end{tabular}
    \caption{\textbf{Reference sample} and \textbf{\ch{N2} samples} - amount of molecular nitrogen added to the reference sample.}
    \label{tab:reference+nitrogen}
\end{table}

\begin{table}
    \setlength{\tabcolsep}{3pt}
    \centering
    \begin{tabular}{||cccccc||}
    \hline
    \multicolumn{6}{||c||}{\textbf{100x solar metallicity}} \\
    \hline\hline
    \textbf{\ch{H2}} & \textbf{\ch{H2O}} & \textbf{\ch{CH4}} & \textbf{\ch{NH3}} & \textbf{\ch{H2S}} & \textbf{\ch{N2}} \\
    \hline
    81.637\%  &  10.470\%   &   6.165\%  & 1.445\% & 0.282\% & 0.428 ppb \\
    \hline
    \end{tabular}
    \caption{\textbf{100x solar metallicity samples} - elemental abundances are chosen to fit chemical equilibrium at the surface and are calculated using \href{https://github.com/exoclime/FastChem}{FastChem} (\protect\cite{2022MNRAS.517.4070S}).}
    \label{tab:100x_solar}
\end{table}

\subsection{Calculating mean mixing ratio}\label{subsec:mix_rat_calc}

To calculate the mean mixing ratio for a specific molecule, number density of air is used and integrated over height, since height in ARGO is increasing in even increments, unlike pressure. Averaging over number density instead of the mixing ratio is chosen, because there is less molecules in the upper atmosphere - if the average was done over the mixing ratio, the result would be skewed towards the higher quantities. 
\protect\begin{equation}
    \langle \chi_M\rangle = \frac{ \int  {\chi_M(h) n(h) \,dh} }{\int {n(h) \,dh}},
\end{equation}
where $\chi_M (h)$ is the mixing ratio of a specie at height $h$ and $n(h)$ is the number density of air at height $h$. Both integrals can be done over the whole atmosphere or its fraction. In the calculations, the mean mixing ratios are obtained for the entire atmosphere, but also for the JWST accessible range only, as these are the values we would expect to detect.
\subsection{Optical depth and stellar flux}\label{subsec:opt-depth}

\begin{figure}
    \centering
    \includegraphics[width=\linewidth]{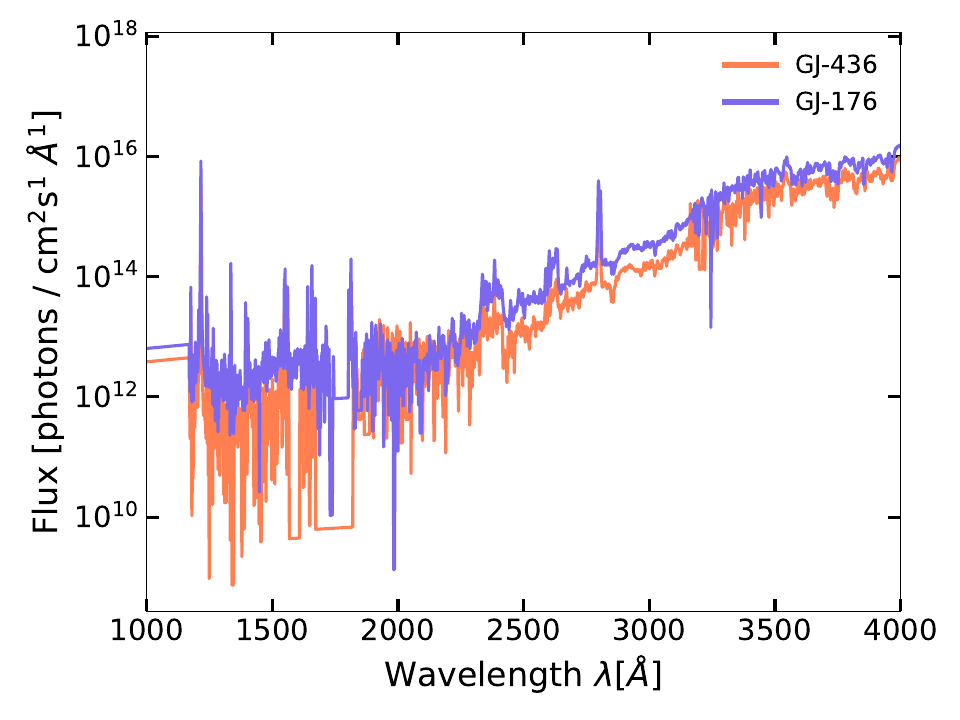}
    \caption{\textbf{Stellar flux} of GJ-176 (violet) and GJ-436 (orange) at different wavelengths in the UV range. Data taken from \protect\href{https://archive.stsci.edu/prepds/muscles/}{MUSCLES Treasury Survey} (\protect\cite{muscles1}, \protect\cite{muscles2}, \protect\cite{muscles3}). The spectrum of GJ-176 was used for all measurements. 
    \label{fig:stellar-flux}}
\end{figure}

\begin{figure}
    \centering
    \includegraphics[width=\linewidth]{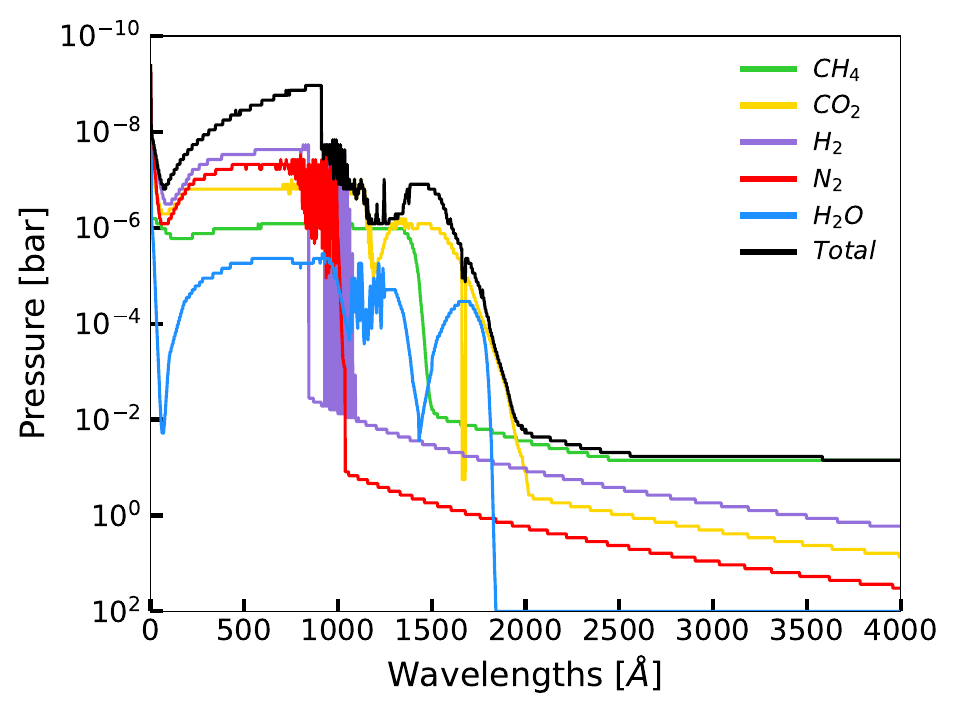}
    \caption{Pressure (bar) at which optical depth, $\tau(\lambda) = 1$, as a function of wavelength ($\text{\AA}$). For the atmosphere described in Table \ref{tab:reference+nitrogen} with $\sim$1\% \ch{N2} (sample D). In the Appendix \ref{Appendix_1}, there is a reference sample with GJ-436 flux for comparison (Figure \ref{fig:ref_stars}).}
    \label{fig:opt-depth}
\end{figure}

The choice of a star is very important - M stars are very active and their spectrum can be variable with time - \cite{Barclay_2021} showed that stellar spots in approximately 1\% of cases, reproduced the same results as atmospheric retrieval of water vapor. Since the full spectrum of K2-18b's host star is not available, all models are run with spectral data of a star of the same type. Model runs were done with UV flux of GJ-176, which was also used by \cite{Wogan_2024} and \cite{Scheucher_2020}. We later compared it with GJ-436 (see Figure \ref{fig:stellar-flux}). These spectra are very similar, however the small differences have an impact on the overall strength of photon absorption by different molecules and optical depth in the higher parts of the atmosphere (see Appendix \ref{Appendix_1} Figure \ref{fig:ref_stars}). 
We would expect these differences to contribute - to see the comparison, see Figure \ref{fig:ref_stars} in Appendix \ref{Appendix_1}. To check the extent to which high energy radiation penetrates the atmosphere and test at what wavelengths certain species absorb the most, one can make use of The Beer-Lambert law for the intensity of light at a given height:
\protect\begin{equation}
    I_{\lambda , z} = I_{\lambda, \infty} \exp{\Big(-\tau_{\lambda}\Big)},
\end{equation}
\protect\begin{equation}
    \tau_{\lambda, z} = \int_{z}^{TOA} \frac{1}{\cos\theta} \sum_{i} \sigma_{\lambda, i} \chi_i (z) n(z) \, dz\ 
\end{equation}
where $\tau$ is the optical depth of the air column and represents absorption of light in the atmosphere (calculated from its top - $TOA$ to the height $z$) due to the number of certain molecules ($ \chi_i (z) n(z)$) and their cross sections ($\sigma_{\lambda, i}$) for reactions with light. $\theta$ is the zenith angle, which in the model is approximately set to 45$^{\circ}$. Large optical depth means that light gets transmitted far into the atmosphere, while small optical depth means that the atmosphere is very hazy and absorbs light quickly, leaving less UV photons to interact with molecules closer to the surface. Figure (\ref{fig:opt-depth}) shows the height at which $\tau_{\lambda} = 1$ for different wavelengths ($\lambda$). Contributions of \ch{H2}, \ch{CH4}, \ch{CO2}, \ch{H2O} and \ch{N2} are also shown. The optical window for \ch{CO2} at around 16$\text{\AA}$ is an artifact of interpolating the cross-section across multiple datasets. This artifact does not affect any of our results.

\subsection{Temperature profile}\label{subsec:pT-profile}

The choice of a temperature profile will have a large impact on the result of this project \citep{Drummond_2016}. It will determine whether thermochemistry plays any role, to what altitude photochemistry dominates and what surface conditions are allowed.  In other studies, simple models assuming partially isothermal atmosphere (\cite{Wogan_2024}) were used, as well as more complicated models, incorporating the effects of clouds, change in water phase or haze (\cite{Piette_2020}). In the literature, there are various approaches to this topic with K2-18b's temperature profiles ranging from colder and habitable to hotter and inhabitable. \cite{Innes_2023} claim that due to \ch{H2} collisional induced absorption, temperatures allowing for liquid water ocean are impossible. However \cite{Madhusudhan_2021} argues that due to the presence of hazes, overall albedo of the planet increases and hence its temperature decreases. 

This study adapts Madhusudhan's approach. The p-T profile was isothermally extrapolated to higher altitudes.
\begin{figure}
    \centering
    \includegraphics[width=\linewidth]{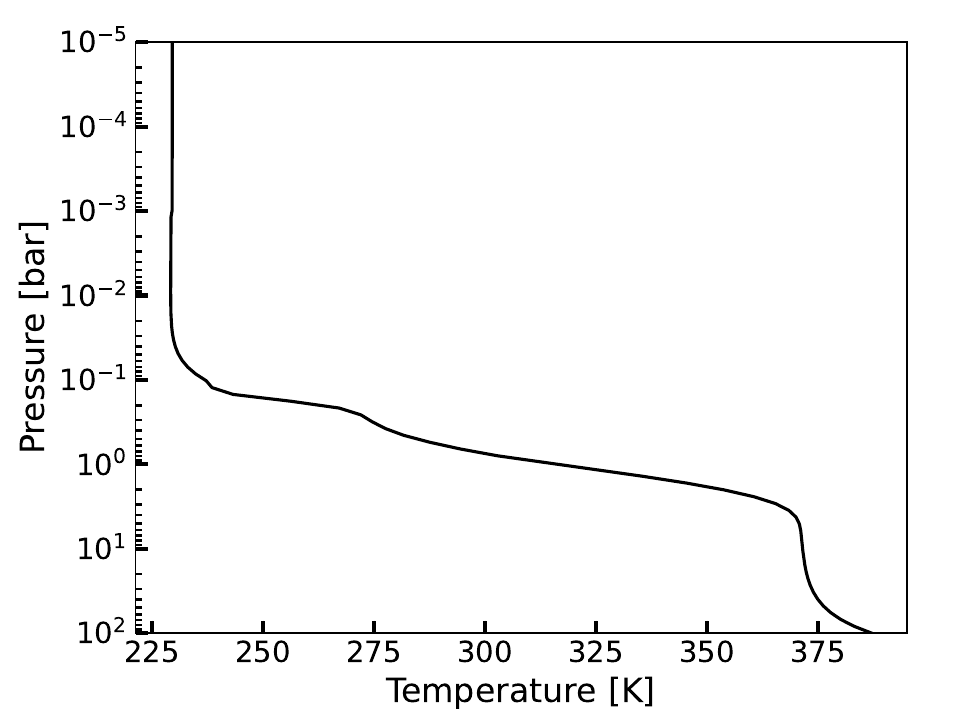}
    \caption{Pressure [bar] against temperature [K] on K2-18b (courtesy of Prof. N. Madhusudhan). Isothermal extrapolation of this profile starts from 10$^{-5}$ bar and continues until 10$^{-10}$ bar. }
    \label{fig:pT-plot}
\end{figure}

\subsection{Eddy Diffusion}\label{subsec:Kzz}

\begin{figure}
    \centering
    \includegraphics[width=\linewidth]{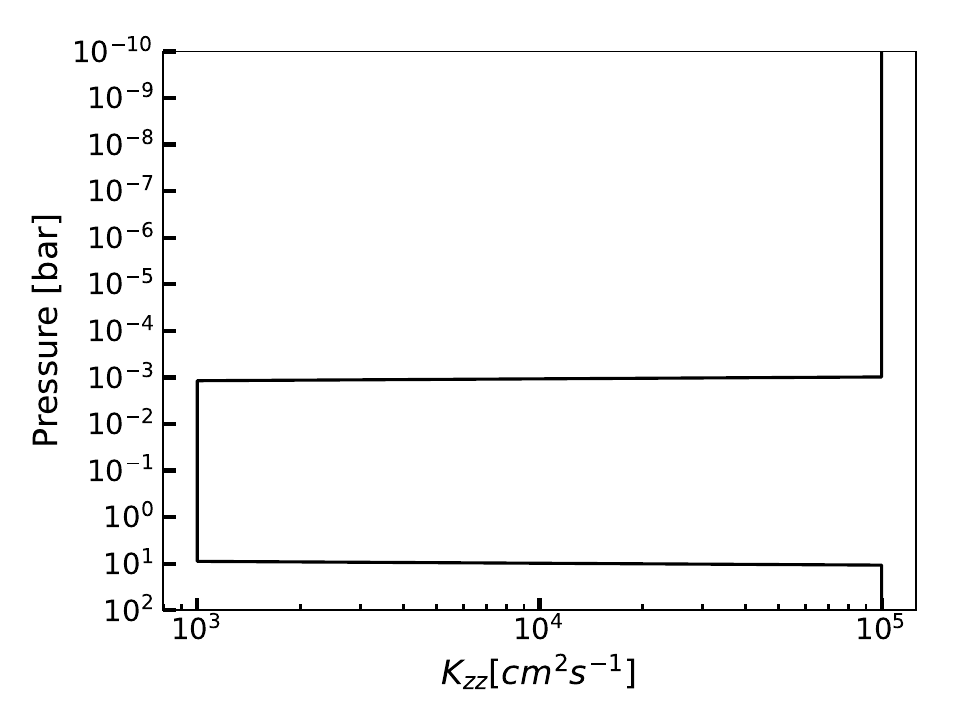}
    \caption{\textbf{Eddy Diffusion profile} - our model of the Kzz trap with $K_{zz} = 10^3$ \unit{cm^2s^{-1}} inside of it and $K_{zz} = 10^5$ \unit{cm^2s^{-1}} throughout the rest of the atmosphere.} 
    \label{fig:Kzz}
\end{figure}

Mixing through Eddy Diffusion is the main way for particle transport in the atmosphere, as molecular diffusion is 5-10 orders of magnitude slower, therefore choosing the correct diffusion parameter ($K_{zz}$) is crucial. There were different approaches in literature regarding K2-18b - some models consisted of constant $K_{zz}$ throughout the whole atmosphere (\cite{Blain_2021}), others included a region with $K_{zz}$ minimum in water condensation region (\cite{Yu_2021}, \cite{Wogan_2024}), which we would call here a 'Kzz trap'. While \cite{Innes_2023} implemented a model for the changing value of $K_{zz}$(T) from \cite{Charnay_2018}. Here we adopt constant $K_{zz}$ profile and one with a Kzz trap (Figure \ref{fig:Kzz}). 

The atmospheric scale height (eq. \ref{eq:H}) of K2-18b is found to be 74.38 \unit{km} (much greater than Earth's atmospheric scale height - 8.5 \unit{km}, because of the difference in molecular weights of hydrogen and nitrogen).
\begin{equation}\label{eq:H}
    H = \frac{RT}{\mu g},
\end{equation}
where $R$ is the gas constant, $T$ is the mean temperature of the atmosphere, $\mu$ the mean molecular weight of the atmosphere and $g$ is the strength of gravitational field of the planet.
For simplicity, mean molecular mass of Jupiter's atmosphere was used ($\mu = 2.33$ \unit{g. mol^{-1}}), as both planets are \ch{H2} dominated. The scale height will vary when changing the composition and its variation is calculated with each run of the code. 
In the lower atmosphere, placing the value of $K_{zz}$ around $10^5 - 10^6$ \unit{cm^2 s^{-1}} agrees well with the estimation from \cite{Charnay_2018}, however following their approach - $K_{zz}$ should increase rapidly in the upper atmosphere, which could impact our photochemistry reaction rates. We have compared our results for constant $K_{zz}$ ranging from $10^5$ \unit{cm^2s^{-1}} to $10^9$ \unit{cm^2s^{-1}}, $K_{zz}$ profile with a "$K_{zz}$ trap" and a $K_{zz}$ profile with Eddy Diffusion coefficient increasing with height (see Appendix \ref{Appendix_1}). Testing more detailed models of Eddy Diffusion was beyond the scope of this project; however, it is a very interesting parameter to explore in future research.

%% file: Results.tex
\section{Results}\label{sec:results}

This section contains the results of simulations and explanations of the chemical reactions leading to production of nitrogen species in K2-18b's atmosphere. It is divided into four subsections describing separate sets of initial conditions: the reference sample \ref{subsec: reference}, the comparison between molecular nitrogen and ammonia (section \ref{subsec:N2 vs NH3}) with \ch{N2} samples \ref{subsec:n2} and 100x solar metallicity sample \ref{subsec: 100x_solar}. At last, a sample with \ch{H2S} \ref{subsec:h2s}, showing significant production of sulfur hazes. The comparison between dry and wet samples is shown in Appendix \ref{Appendix_5}. Detailed discussion of the findings in the context of literature is shown in section \ref{sec:discussion}. The chemical mechanisms in sub-Neptunes (both cold and hot) were extensively studied in the past and it is instructive to compare our results with the literature (\cite{Yung_1984}, \cite{Lavvas_2008}, \cite{Line_2011}, \cite{Moses_2011}, \cite{Venot_2012}, \cite{Agundez_2014}, \cite{KRASNOPOLSKY_2014}, \cite{VUITTON_2019}, \cite{Mollier_2015}, \cite{Moses_2016}, \cite{Blumenthal_2018}, \cite{Kawashima_2019}, \cite{Hobbs_2019}, \cite{Lavvas_2019}, \cite{Molaverdikhani_2019}, \cite{Hu_2021}, \cite{Rimmer_2019}).

\subsection{Reference sample}\label{subsec: reference}
The reference sample consists of only 3 inputs: \ch{H2}, \ch{CH4} and \ch{CO2} (see Table \ref{tab:reference_sample}). It was tested among all of the Eddy Diffusion profiles (Figure \ref{fig:Kzz}) and with the flux of both stars (comparison between the stars was done with $K_{zz}=10^6$ \unit{cm^2 s^{-1}}) (Figure \ref{fig:ref_stars}). The shape of height profiles for corresponding molecules is very similar for lower $K_{zz}$ values ($K_{zz}=10^4 - 10^6$ \unit{cm^2 s^{-1}}), but differs for $K_{zz} \geq 10^7$ \unit{cm^2 s^{-1}}. The distinction between slow and fast diffusion can be made, because somewhere between $K_{zz}=10^6 - 10^7$ \unit{cm^2 s^{-1}}, the rate of chemical reactions becomes smaller than the rate of transport and hence the process determining the lifetime of a specie at a given height changes. It is reflected in the shape of the profile - comparison between slow and fast diffusion is presented on Figure \ref{fig:reference_Kzz} in Appendix \ref{Appendix_1}. Because of the domination of transport at higher Eddy Diffusion values, to uncover the interesting chemistry of K2-18b's atmosphere,  most of the simulations were done for $K_{zz}$ between $10^5$ and $10^6$ \unit{cm^2 s^{-1}}.

\begin{figure}
\centering
    \includegraphics[width=\linewidth]{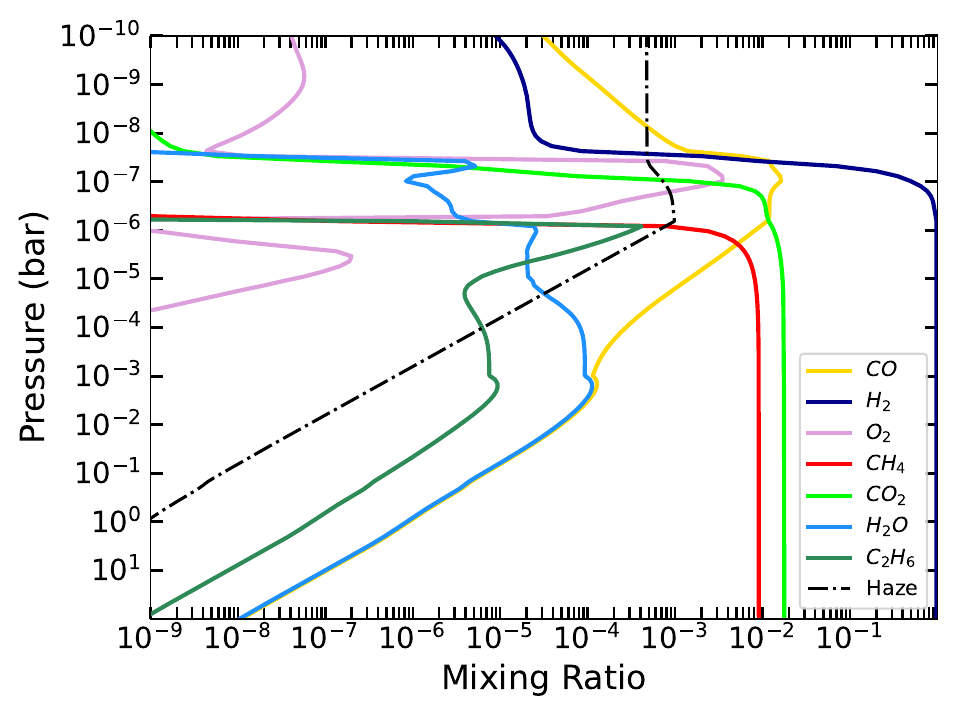}
    \caption{Height profile for $K_{zz}$ trap reference sample - initially, there were only 3 molecules present - \ch{H2}, \ch{CO2}, \ch{CH4}, the rest was created through chemical reactions from \textbf{Stand2023} (\protect\cite{Rimmer_2016}). The 'bump' around p $=10^{-5}$ bar is due to Eddy Diffusion profile for the $K_{zz}$ trap sample - a step change in $K_{zz}$ value from $10^3$ in the $K_{zz}$ trap to $10^5$ \unit{cm^2.s^{-1}} in the rest of the atmosphere.}
    \label{fig:N2_0_cold trap}
\end{figure}

The assumption of hydrogen rich atmosphere combined with substantial amounts of methane detected, results in a large reservoir of C and H atoms and in significant production of hydrocarbons in all of the tested samples. Larger hydrocarbons do not participate in the more interesting reactions, hence they are plotted together as \textbf{Haze} (see  Table \ref{tab:Haze}). \cite{Lavvas_2019} showed that increasing metallicity increases haze production, however due to large abundance of methane, photochemical hazes are produced in all of our samples. More active hydrocarbons are: \ch{C2H2}, \ch{C2H4} and \ch{C2H6} - they play a major role in \ch{HC3N} production in samples with nitrogen. 
\begin{figure}
    \centering
    \includegraphics[width=\linewidth]{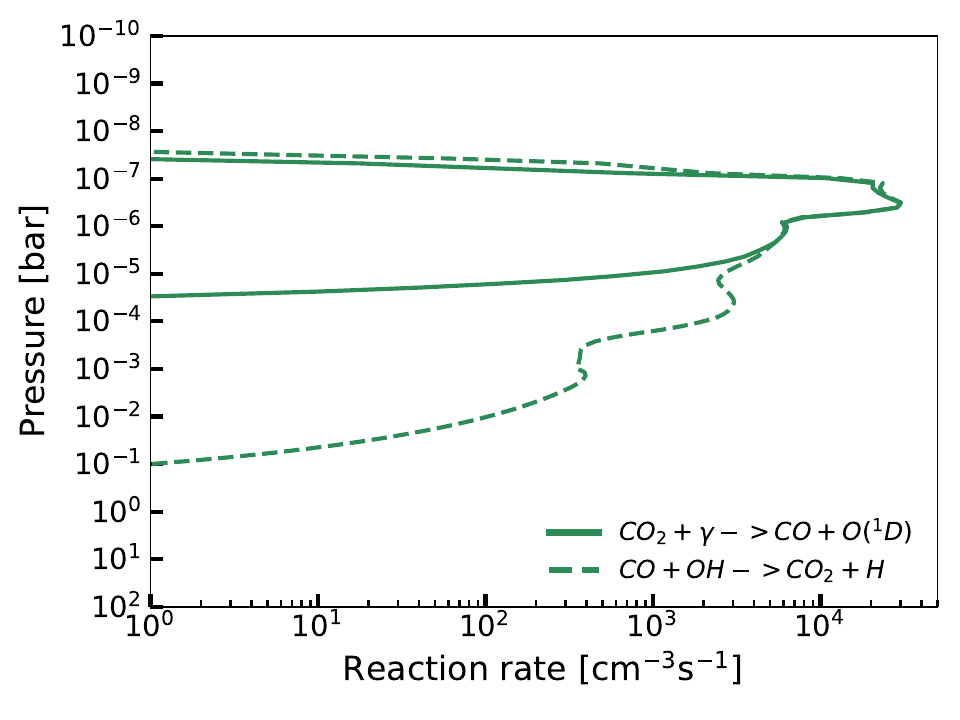}
    \caption{Reaction rates for \textbf{CO} production through \ch{CO2 + $h\nu$} $\rightarrow$ \ch{CO + O(^1D)} and destruction through \ch{CO + OH} $\rightarrow$ \ch{CO2 + H} in the reference sample with $K_{zz}$ trap.}
    \label{fig:CO_rate}
\end{figure}
The production of water, \ch{CO} and ethane (\ch{C2H6}) is common for all samples, regardless of their Eddy Diffusion regime or additional inputs.
The production of CO through \ch{CH4}-CO conversion is less important here than in \cite{Moses_2011} and \cite{Moses_2016}, because of the high abundance of \ch{CO2}. In the upper atmosphere, more energetic photons break \ch{CO2} molecules apart to form \ch{CO} and reactive oxygen atoms \ch{O($^1$D)}, which quickly loose their energy on collisions with \ch{H2} molecules. Effectively, this process can be described by a net reaction (see Figure \ref{fig:CO_rate}): \\

\ch{CO2 + $h\nu$} $\rightarrow$ \ch{CO + O}.\\  

\begin{figure}
    \centering
    \includegraphics[width=\linewidth]{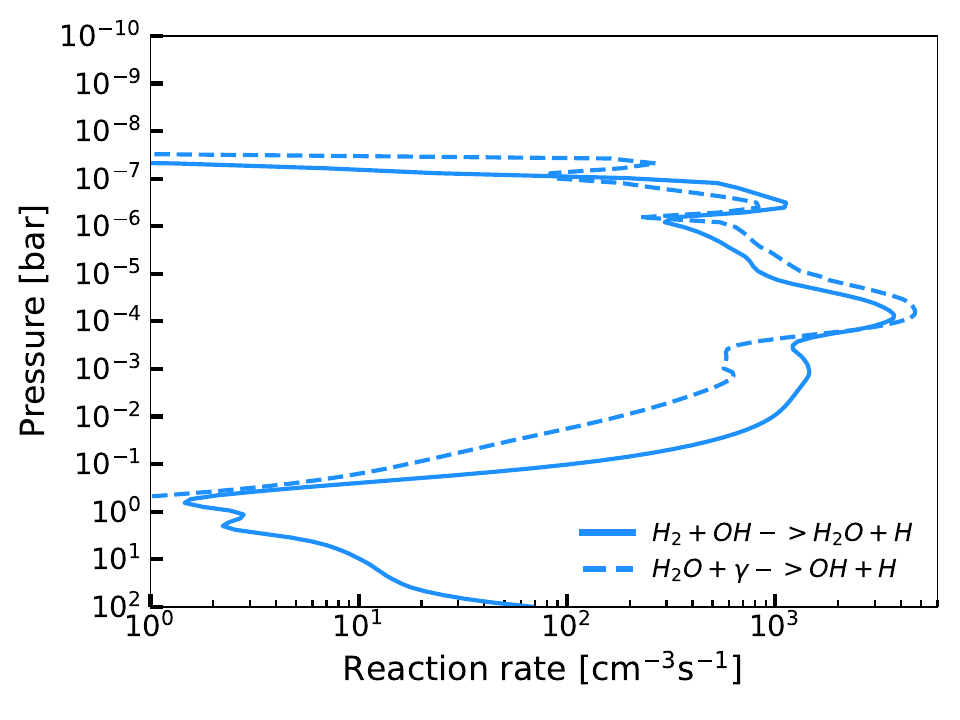}
    \caption{Reaction rates for \textbf{\ch{H2O}} production through \ch{H2 + OH } $\rightarrow$ \ch{ H2O + H} (solid line) and destruction through \ch{H2O + $h\nu$ } $\rightarrow$ \ch{ OH + H} (dashed line) in the reference sample with $K_{zz}$ trap.}
    \label{fig:water_rate}
\end{figure}

However, \ch{CO2} molecules in the deeper atmosphere follow a different path. Oxygen produced from the photolysis of carbon dioxide combines with \ch{H2} to produce \ch{OH} radicals and consequently - water. The two reactions below, are responsible for the production of \ch{CO} and \ch{OH} in the atmosphere, which later facilitate the destruction of methane and production of \ch{NO}. The whole process results in simultaneous creation of \ch{CO} and \ch{H2O} molecules (see Figure \ref{fig:water_rate}):\\

\hspace{-0.4cm} \ch{CO2 + $h\nu$} $\rightarrow$ \ch{CO + O($^1$D)}\\
\ch{H2 + O($^1$D)} $\rightarrow$ \ch{OH + H}\\ 
\ch{H2 + OH} $\rightarrow$ \ch{H2O + H}\\
\ch{H + H} $\rightarrow$ \ch{H2}\\ 
\textbf{Net: }\ch{CO2 + H2 + $h\nu$} $\rightarrow$ \ch{H2O + CO}.\\

This result is important to estimate the amount of water or carbon monoxide in the atmosphere for the dry case scenario, should any of these molecules be detected in the future. CO production in a deep atmosphere is slightly different for the wet scenario - see Appendix \ref{Appendix_5} for details.
It can be seen that all of these chemical processes are strongly related to each other.
The production of all hydrocarbons starts with the destruction of methane (unlike in \cite{Moses_2016}, where water and \ch{C2H6} take part in acetylene production). Acetylene (\ch{C2H2}) and polyethylene (\ch{C2H4}) follow the same path:\\

\hspace{-0.4cm} \ch{CH4 + $h\nu$} $\rightarrow$ \ch{CH2 + 2 H} \\
\ch{CH2 + H} $\rightarrow$ \ch{CH + H2} \\
\ch{CH4 + CH} $\rightarrow$ \ch{C2H4 + H} \\
\ch{C2H4 + $h\nu$} $\rightarrow$ \ch{C2H2 + H2}\\
\ch{H + H} $\rightarrow$ \ch{H2}\\
\textbf{Net:} \ch{2 CH4 + 2 $h\nu$} $\rightarrow$ \ch{C2H2 + 3 H2}.\\

\hspace{-0.4cm} \ch{CH4 + $h\nu$} $\rightarrow$ \ch{CH2 + 2 H} \\
\ch{CH2 + H} $\rightarrow$ \ch{CH + H2} \\
\ch{CH4 + CH} $\rightarrow$ \ch{C2H4 + H} \\
\ch{H + H} $\rightarrow$ \ch{H2}\\
\textbf{Net:} \ch{2 CH4 + $h\nu$} $\rightarrow$ \ch{C2H4 + 2 H2}.\\

\begin{figure}
    \centering
    \includegraphics[width=\linewidth]{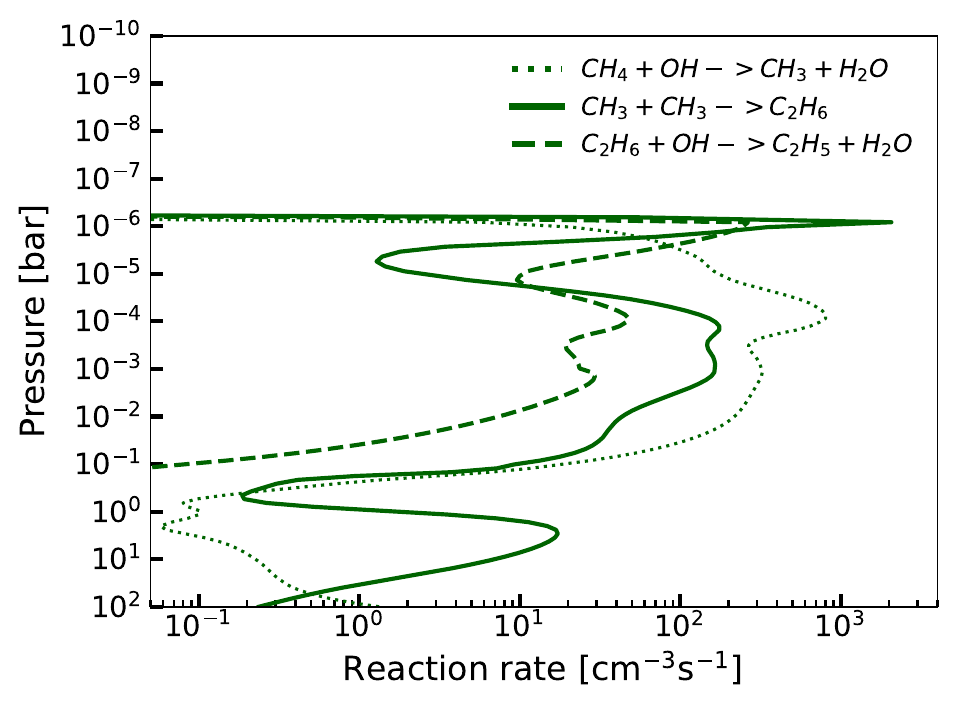}
    \caption{Reaction rates for the reference sample with $K_{zz}$ trap for \textbf{\ch{C2H6}} production through \ch{CH3 + CH3 } $\rightarrow$ \ch{ C2H6} (solid) and destruction through \ch{C2H6 + OH} $\rightarrow$ \ch{C2H5 + H2O} (dashed). Additionally there is plotted a reaction rate for the production of methyl radicals (\ch{CH3}): \ch{CH4 + OH } $\rightarrow$ \ch{ CH3 + H2O} (dotted).}
    \label{fig:ethane_rate}
\end{figure}

\ch{OH} radicals combine with methane to produce methyl radicals - \ch{CH3}, which later collide to form ethane (see Figure \ref{fig:ethane_rate}):\\

\hspace{-0.4cm} 2 x (\ch{CO2 + $h\nu$} $\rightarrow$ \ch{CO + O($^1$D)})\\
2 x (\ch{CH4 + OH} $\rightarrow$ \ch{CH3 + H2O})\\
2 x (\ch{H2 + O($^1$D)} $\rightarrow$ \ch{OH + H})\\
\ch{CH3 + CH3} $\rightarrow$ \ch{C2H6}\\
\textbf{Net:} \ch{2 CO2 + 2 CH4 + H2 + 2 $h\nu$} $\rightarrow$ \ch{C2H6 + 2 CO + 2 H2O}.\\

 The peak production of \ch{C2H6} coincides with the peak production of \ch{H2O} and \ch{CO}. The appearance of ethane in significant quantities can have some interesting implications, which will be discussed further in Section \ref{sec:discussion}. As can be seen in Figure \ref{fig:ethane_rate}, the production of methyl radicals depends directly on the amount of methane present in the atmosphere: the peak in \ch{C2H6} disappears immediately as soon as the methane is depleted around the $10^{-6}$ bar (see Figure \ref{fig:N2_0_cold trap}). In the wet scenario, the OH radicals come from the dissociation of water (see Appendix \ref{Appendix_5}). Since we know well the amount of methane present (\cite{Madhusudhan_2023}), any surplus production of \ch{CH3} should be noticed through increased amounts of ethane. These extra radicals could come from the destruction of dimethyl sulfide as suggested by \cite{Domagal_2011}.

\subsection{\ch{N2} vs \ch{NH3}}\label{subsec:N2 vs NH3}
\begin{figure*}
    \centering
    \begin{subfigure}[b]{0.48\linewidth}
        \includegraphics[width=\linewidth]{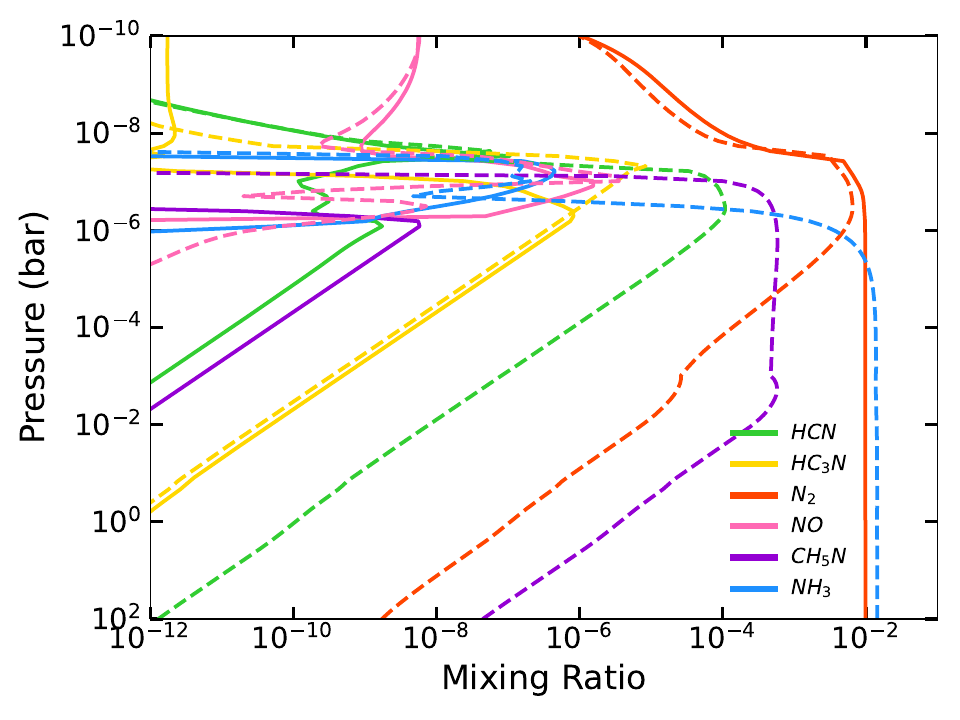}
        \caption{Comparison of height profiles of nitrogen compounds in a sample with $\sim$ 1\% of added \ch{N2} (solid lines) and a sample with 100x solar metallicty (dashed lines) with ammonia as a dominant source of nitrogen in the atmosphere.}
        \label{fig:height}
    \end{subfigure}
    \hfill
    \begin{subfigure}[b]{0.48\linewidth}
        \includegraphics[width=\linewidth]{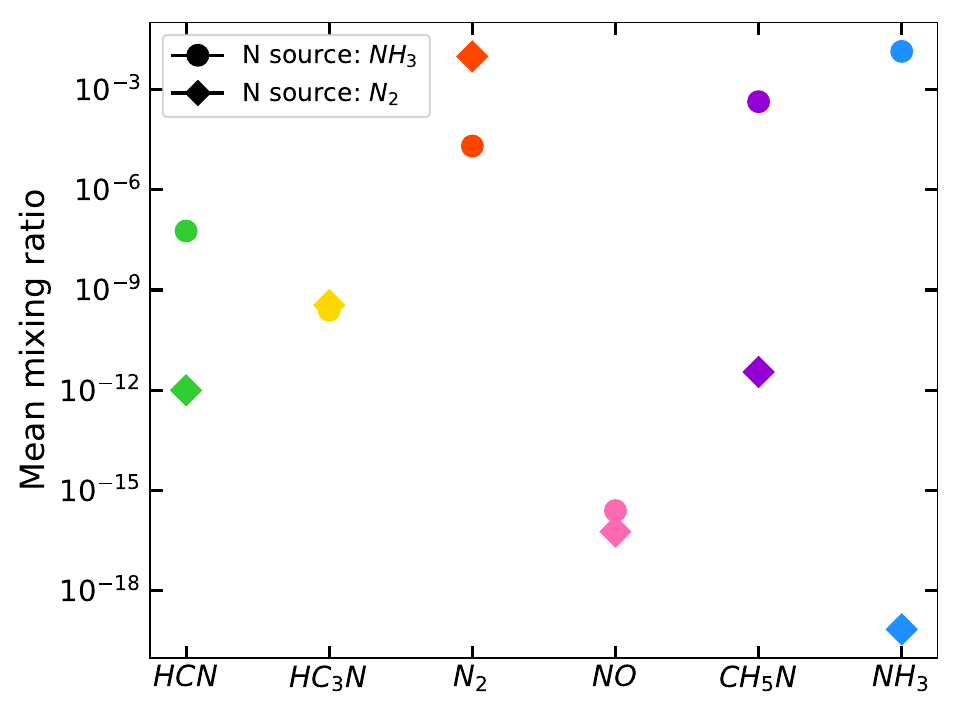}
        \caption{Plot of mean mixing ratio of nitrogen compounds in the detectable region ($10^{-5}$ to $10^{-2}$ bar) - comparison of $\sim$1\% added \ch{N2} (sample D) and 100x solar metallicity (\ch{NH3} as a dominant nitrogen source).}
        \label{fig:summary}
    \end{subfigure}
    \caption{Comparison of samples with nitrogen sources dominated by \ch{N2} and \ch{NH3}. On the left panel, there are height profiles for 6 different nitrogen species: \ch{NH3}, \ch{HCN}, \ch{HC3N}, \ch{N2}, \ch{NO} and \ch{CH5N}. On the right panel, there are mean mixing ratios of these compounds in the detectable region. One can see, that apart from \ch{HC3N} and \ch{NO}, the differences between samples are major (mind - logarithmic scale on the y-axis). It suggests that these 2 compounds are bad proxies for differentiating between the sources of nitrogen in the atmosphere, while all the others could be used to differentiate between \ch{N2} and \ch{NH3} nitrogen sources in the exoplanetary atmospheres.}
    \label{fig:summary_plots}
\end{figure*}

The differences between model outputs for \ch{N2} based and \ch{NH3} based scenarios are significant (see Figure \ref{fig:summary_plots}). Break up of \ch{N2} molecules through photolysis results overall in lower production of other nitrogen compounds compared to the case with \ch{NH3}. Additionally, their production starts higher in the atmosphere, which reduces our chances of detecting them in the future. Section \ref{subsec:n2} provides the details of \ch{N2} chemistry and section \ref{subsec: 100x_solar} describes the sample with the domination of ammonia. 

\subsubsection{Samples with \ch{N2} from 10 ppm to 10\%}\label{subsec:n2}

\begin{figure*}
    \includegraphics[width=\linewidth]{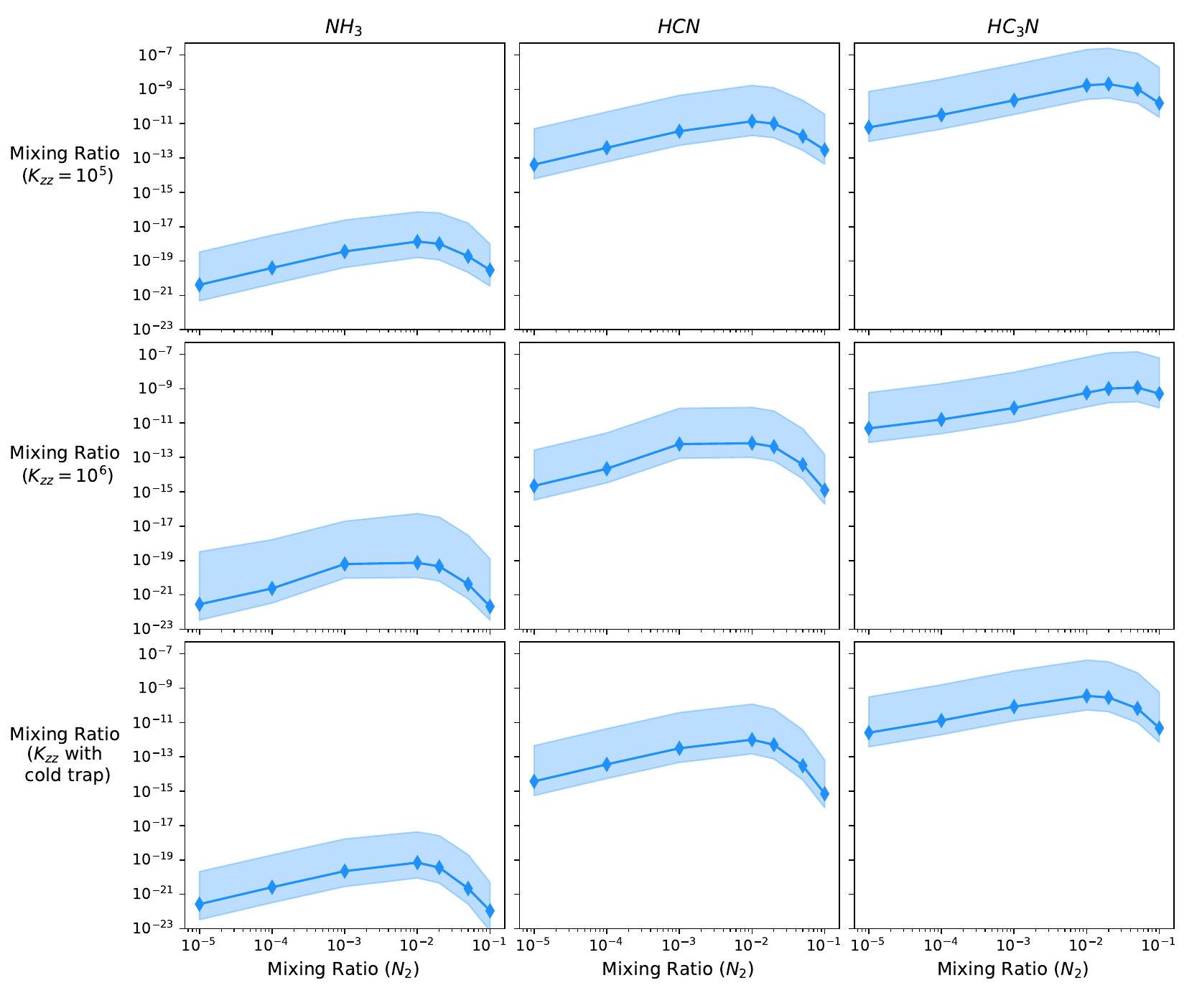}
    \caption{Mixing ratio of N bearing compounds versus \ch{N2} input for samples with lower Eddy Diffusion coefficients (from the top: $K_{zz}=10^5$ \unit{cm^2 s^{-1}}, $K_{zz}=10^6$ \unit{cm^2 s^{-1}} and $K_{zz}$ with '$K_{zz}$ trap'). Calculations are restricted to the detectable range of the spectrum ($10^{-5}$ to $10^{-2}$ bar). Diamond pointers represent \ch{N2} input values and shaded region is in between minimum and maximum mixing ratios of a given specie in the region of interest. Varying $K_{zz}$ has little to no effect on the mean mixing ratio of a specie, however, the values of maximum and minimum amounts of it in the region change - you can see that through the variation of shaded region's width. For larger amounts of added nitrogen, the shape of the curve varies slightly with $K_{zz}$ - it flattens or deepens depending on the specie. None of the nitrogen compounds is produced in large enough quantities (mixing ratio $> 10^{-5}$) to be detected.}
    \label{fig:N2_plots}
\end{figure*}

\begin{figure}
    \centering
    \includegraphics[width=\linewidth]{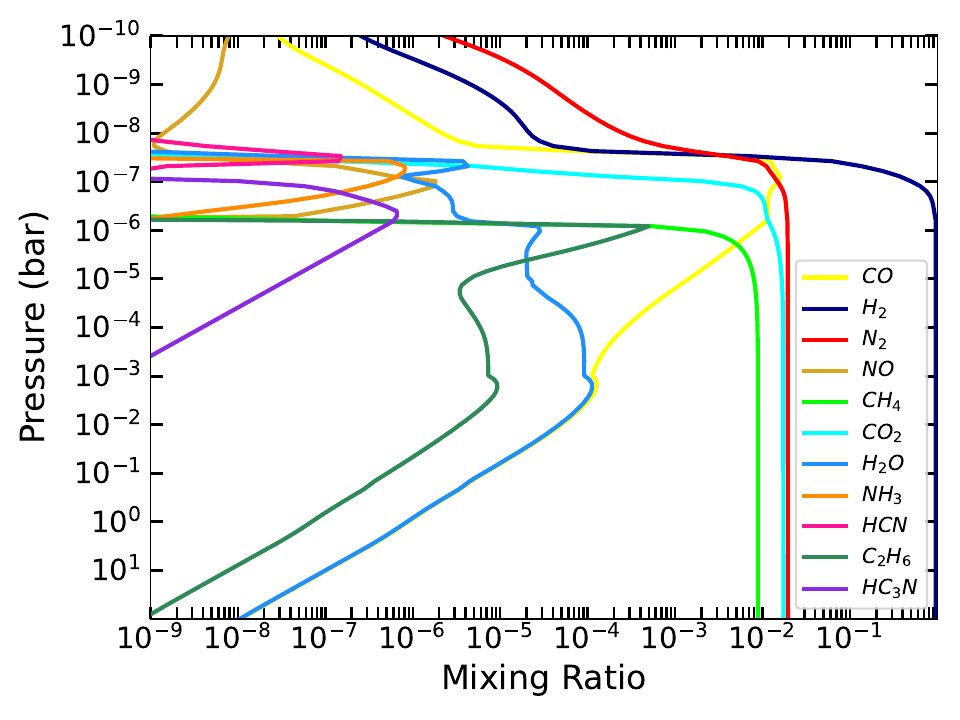}
    \caption{Height profile for $K_{zz}$ trap sample with 2\% of \ch{N2} added. For clarity, there are only plotted molecules taking part in the production of nitrogen bearing species, however all the inputs and products from Figure \ref{fig:N2_0_cold trap} are also present.}
    \label{fig:N2.2p_cold-trap}
\end{figure}

As expected, molecular nitrogen is fairly inert in the atmosphere - it has no effect on the surface chemistry and only becomes active in the top parts of the atmosphere, where photochemistry initiates its further reactions. The addition of \ch{N2} to the atmosphere has little effect on the shape of height profiles of species in the reference sample (compare Figures \ref{fig:N2_0_cold trap} and \ref{fig:N2.2p_cold-trap}). The only exception is the rate of depletion of \ch{CO} at around p $= 10^{-8}$ bar. It increases significantly, when nitrogen is present in the atmosphere. The reason behind it is a reaction between \ch{CO} and \ch{N2+} ions: 
\ch{CO + N2+} $\rightarrow$ \ch{CO+ + N2}. \\

For all samples, the peak production of nitrogen compounds is situated between 10$^{-6}$ and 10$^{-8}$ bar. Around that pressure, photodissociation of \ch{N2} results in the making of a few important species: \ch{NH3}, \ch{HCN}, \ch{NO} and \ch{HC3N}.
Ammonia production is initiated when \ch{N+} reacts with molecular hydrogen and follows the same mechanism as described in \cite{Stark_Helling_Diver_Rimmer_2014}, a chemistry that carries an interstellar flavour:\\

\hspace{-0.4cm} 2 x (\ch{N2 + $h\nu$} $\rightarrow$ \ch{N+ + N + e-})\\
2 x (\ch{N+ + H2 } $\rightarrow$ \ch{ NH+ + H})\\
2 x (\ch{NH+ + H2 } $\rightarrow$ \ch{ NH2+ + H})        \\
2 x (\ch{NH2+ + H2 } $\rightarrow$ \ch{ NH3+ + H})        \\
2 x (\ch{NH3+ + H2} $\rightarrow$ \ch{ NH4+ + H})        \\
2 x (\ch{NH4+ + e- } $\rightarrow$ \ch{ NH3 + H})         \\ 
\ch{N + N} $\rightarrow$ \ch{N2} \\
5 x (\ch{H + H} $\rightarrow$ \ch{H2}) \\
\textbf{Net:} \ch{N2 + 3 H2 + 2 $h\nu$} $\rightarrow$ 2 \ch{NH3}\\

However, deeper in the atmosphere, ammonia is produced alongside HCN through the destruction of methane and carbon dioxide:\\

\hspace{-0.4cm} \ch{CH4 + M} $\rightarrow$ \ch{CH3 + H + M}\\
\ch{N2 + H} $\rightarrow$ \ch{HN2}\\
\ch{HN2 + CO2} $\rightarrow$ \ch{NO + HNCO}\\
\ch{NO + CH3} $\rightarrow$ \ch{HCN + H2O}\\
\ch{HNCO + H} $\rightarrow$ \ch{CO + H2N}\\
\ch{H2N + H2} $\rightarrow$ \ch{NH3 + H}\\
\textbf{Net:} \ch{N2 + H2 + CH4 + CO2} $\rightarrow$ \ch{HCN + NH3 + H2O + CO}\\

The fastest way to get cyanoacetylene is through the destruction of hydrogen cyanide:\\

\hspace{-0.4cm} \ch{CH4 + $h\nu$} $\rightarrow$ \ch{CH2 + 2 H}\\
\ch{CH2 + H} $\rightarrow$ \ch{CH + H2}\\
\ch{CH4 + CH} $\rightarrow$ \ch{C2H4 + H}\\
\ch{C2H4 + $h\nu$} $\rightarrow$ \ch{C2H2 + H2}\\
\ch{HCN + $h\nu$} $\rightarrow$ \ch{CN + H}\\
\ch{CN + C2H2} $\rightarrow$ \ch{HC3N + H}\\
2 x \ch{(H + H $\rightarrow$ H2)}\\
\textbf{Net:} \ch{2 CH4 + HCN + 3 $h\nu$} $\rightarrow$ \ch{HC3N + 4 H2}\\

HCN production occurs with the help of \ch{CH3} radicals and follow the same pathway as in \cite{Moses_2016} and \cite{Line_2011}, but here the source of N is \ch{N2} and not \ch{NH3}, catalyzed by \ch{H2O}:\\

\hspace{-0.4cm} \ch{N2 + CH $\rightarrow$ HCN + N}\\
\ch{H2CN + $h\nu$ $\rightarrow$ HCN + H}\\
\ch{N + CH3 $\rightarrow$ H2CN + H}\\
3 x \ch{(H2O + $h\nu$ $\rightarrow$ OH + H)}\\
2 x \ch{(CH4 + OH $\rightarrow$ CH3 + H2O)}\\
\ch{OH + CH3 $\rightarrow$ CH2 + H2O}\\
\ch{CH2 + H $\rightarrow$ CH + H2}\\
2 x \ch{(H + H $\rightarrow$ H2)}\\
\textbf{Net:} \ch{2 CH4 + N2 + 4 $h\nu$ $\rightarrow$ 2 HCN + 3 H2}\\

All of these processes are consistent for all nitrogen samples, except from their reaction rates - more \ch{N2} abundance accelerates the production of some species, which later affects the abundance of other species. The comparison between samples with different amounts of \ch{N2}, but the same $K_{zz}$ profile is shown on Figure \ref{fig:N2_Kzz_6}. The comparison between dry and wet samples with \ch{N2} - their height profiles, total abundances of species and their formation pathways is shown in Appendix \ref{Appendix_5}.

\subsubsection{100x solar metallicity}\label{subsec: 100x_solar}

\begin{figure}
    \centering
    \includegraphics[width=\linewidth]{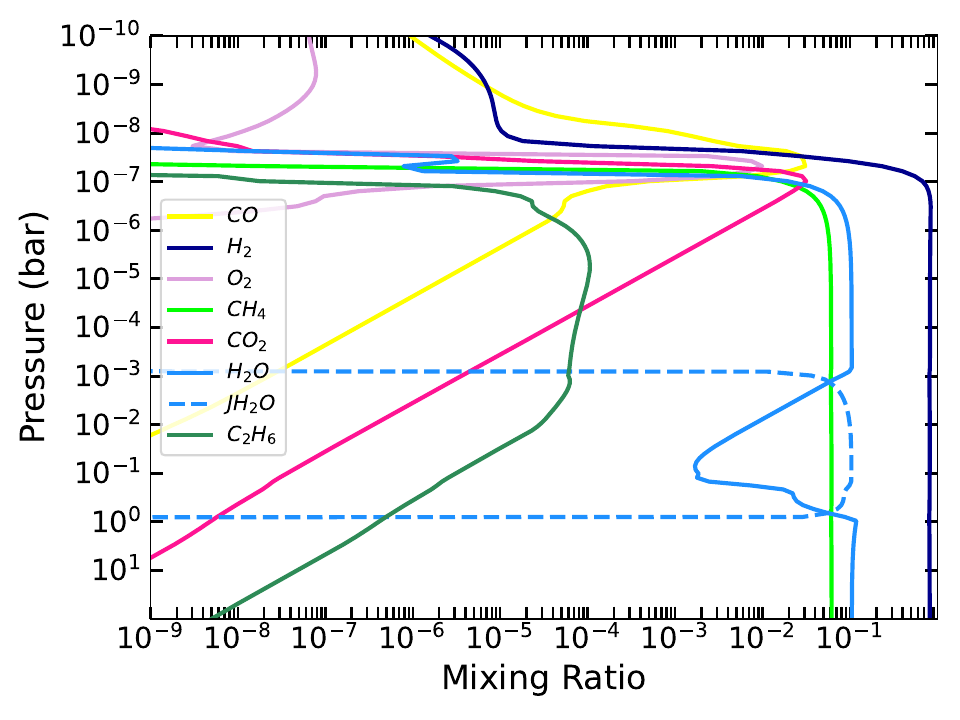}
    \caption{Height profile for $K_{zz}$ trap sample with 100x solar metallicity - carbon, hydrogen and oxygen bearing species. Dashed blue curve (\ch{JH2O}) is the condensed water vapor.}
    \label{fig:100x_carbon-water}
\end{figure}

\begin{figure}
    \centering
    \includegraphics[width=\linewidth]{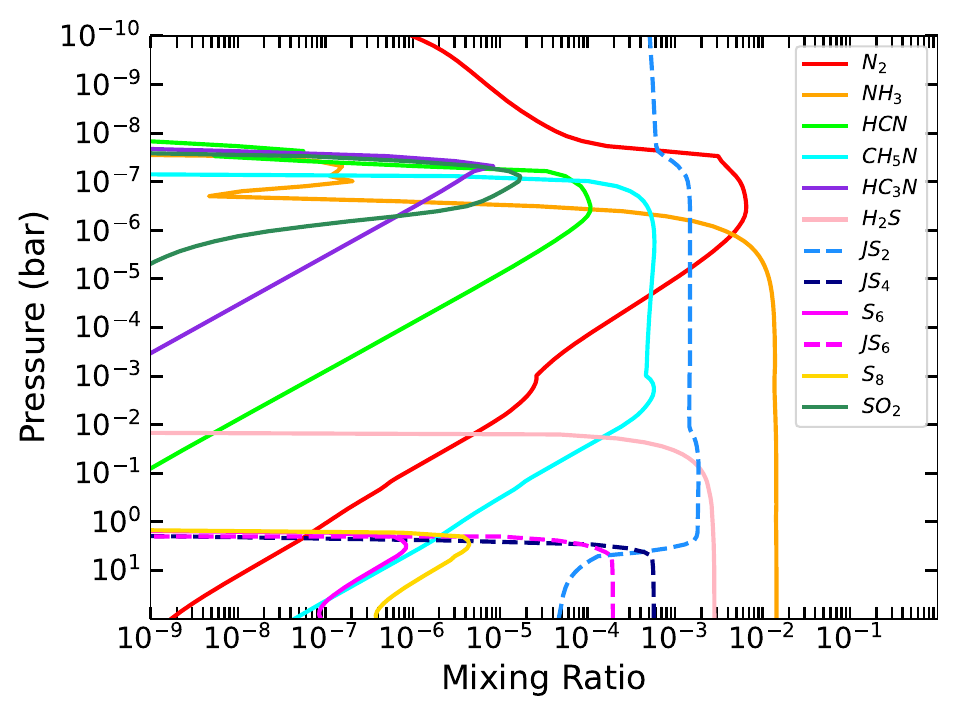}
    \caption{Height profile for $K_{zz}$ trap sample with 100x solar metallicity - nitrogen and sulfur bearing species. 'J' denotes condensed species.}
    \label{fig:100x_nitrogen_sulfur}
\end{figure}

To investigate the chemistry of nitrogen source from ammonia, a sample with 100x solar metallicity (Table \ref{tab:100x_solar}) was tested in different Eddy Diffusion profiles. Initial conditions at the bottom of the atmosphere are dictated by the equilibrium chemistry of species adapted from \cite{Asplund_2021} for 100x solar metallicity. Chemical equilibrium was calculated using \href{https://github.com/exoclime/FastChem}{FastChem} \citep{2022MNRAS.517.4070S}. For clarity, results are divided into 3 different plots: Figure \ref{fig:100x_carbon-water} shows carbon and water chemistry, Figure \ref{fig:100x_nitrogen_sulfur} nitrogen \& sulfur chemistry. Figure \ref{fig:100x_cold trap} represents comparison between equilibrium chemistry (dotted lines) and ARGO output (solid lines). It shows really well the effect of photodissociation of species at the top of the atmosphere (ARGO) and thermal recycling at the bottom (which only happens at equilibrium conditions).

Most of the inputs (like \ch{NH3}, \ch{H2}, \ch{H2O} and \ch{CH4}) are fairly stable throughout the atmosphere and the main process of their destruction is through photodissociation in the UV active region. The exception to that rule are sulphur species: \ch{H2S} is depleted in the $K_{zz}$ trap, due to the reaction with the hydrogen atom released from ammonia destruction, creating \ch{S2} condensate as a result. The breaking apart of the molecules of ammonia is the dominating pathway to production of all nitrogen species in this sample. 

\begin{figure}
    \centering
    \includegraphics[width=\linewidth]{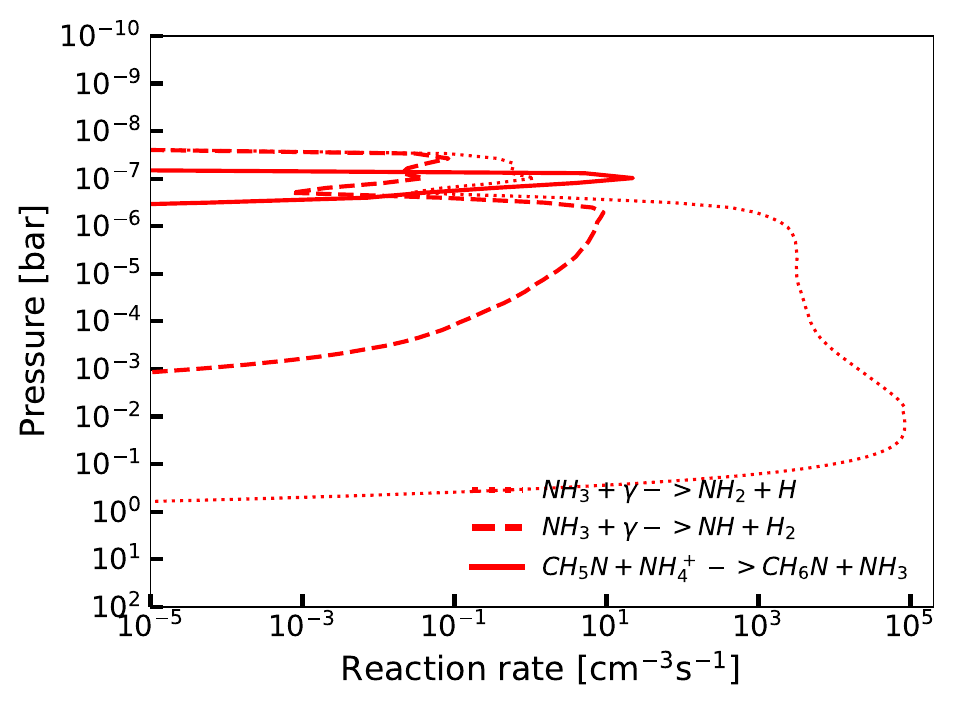}
    \caption{Main pathways for the destruction (dashed and dotted lines) of ammonia in 100x solar metallicity sample with $K_{zz}$ trap. The solid red curve shows the reaction producing \ch{NH3} from the destruction of \ch{CH5N} - reaction dominates at around 10$^{-7}$ bar. It is the only place in the atmosphere, where the production of ammonia dominates over destruction of it. The difference between production and destruction rate is very large - see the logarithmic scale at the bottom.}
    \label{fig:ammonia_rate}
\end{figure}

Peak production for most of nitrogen bearing species (Figure \ref{fig:100x_nitrogen_sulfur}) occurs close to $10^{-6} \sim 10^{-7}$ bar in the atmosphere and coincides with the fastest ammonia destruction. Methylamine (\ch{CH5N}) is produced through a simple reaction, where one hydrogen atom in \ch{NH3} is replaced by a single methyl group. It is unique to this sample - it was not observed in samples with nitrogen coming from \ch{N2}. Ammonia could also break up to a \ch{NH} radical and \ch{H2} molecule and lead to \ch{HCN} production, however the rate of this reaction is much lower, therefore the production of methylamine dominates (see Figure \ref{fig:ammonia_rate}). 

Unlike in the previous samples, here, \ch{CO2} is not an input, but is created together with \ch{CO}, through \ch{CH4} and \ch{OH}:\\

\hspace{-0.4cm} 3 x (\ch{H2O + $h\nu$} $\rightarrow$ \ch{OH + H})\\
\ch{CH4 + OH } $\rightarrow$ \ch{ CH3 + H2O}\\
\ch{CH3 + OH } $\rightarrow$ \ch{ CH2O + H2}         \\
\ch{CH2O + H } $\rightarrow$ \ch{ CHO + H2}        \\
\ch{CHO + H} $\rightarrow$ \ch{ CO + H2}        \\
\ch{CO + OH} $\rightarrow$ \ch{ CO2 + H}         \\ 
\ch{ H + H} $\rightarrow$ \ch{H2} \\
\textbf{Net:} \ch{2 H2O + CH4 } $\rightarrow$ \ch{ CO2 + 4 H2}\\
 
In the upper atmosphere, \ch{CO} formation pathway follows the same reactions and with the addition of \ch{CO2} photolysis, the net result yields two times more \ch{CO} molecules than water: 
\ch{3 CO2 + CH4} $\rightarrow$ \ch{4 CO + 2 H2O}.\\
It is consistent, as \ch{H2O} destruction rate is higher than production at around $10^{-7}$ bar and \ch{CO} production has a peak there. However, for the most of the atmosphere, water is at a much higher abundance than carbon monoxide, which could be another feature differentiating \ch{NH3} rich (equilibrium) from \ch{N2} rich samples. It is also the only sample in which water has high enough partial pressure to condense - height of the formed cloud deck coincides with the placement of the $K_{zz}$ trap in the atmosphere (see Figure \ref{fig:100x_carbon-water}).

Same as in the previous samples, ethane (\ch{C2H6}) is the dominant hydrocarbon. Similarly to the reference sample (\ref{subsec: reference}), it is constructed through the creation of \ch{CH3} radicals from the destruction of methane, however in this case, it happens through a combination with H atoms released from ammonia: \ch{CH4 + H}  $\rightarrow$ \ch{CH3 + H2}. 

\ch{N2} production is the reverse mechanism of \ch{NH3} production from \cite{Moses_2016} and is the same as low pressure model of hot Jupiter-like planet in \cite{Moses_2011}. It originates from ammonia photolysis and follows a pathway through the formation of diimide:\\

\hspace{-0.4cm} 2 x (\ch{NH3 + $h\nu$} $\rightarrow$ \ch{NH2 + H})\\
\ch{NH2 + NH2 } $\rightarrow$ \ch{ HNNH + H2}\\
\ch{HNNH + H } $\rightarrow$ \ch{ N2H + H2}\\
\ch{N2H + $h\nu$ } $\rightarrow$ \ch{ N2 + H}\\ 
\ch{ H + H} $\rightarrow$ \ch{H2} \\
\textbf{Net:} \ch{2 NH3 + 2 $h\nu$} $\rightarrow$ \ch{ N2 + 3 H2}\\

Around the pressure $10^{-7}$ bar, there is an additional peak in ammonia, which comes from the destruction of \ch{N2} and results in a net reaction:
\ch{N2 + CH4} $\rightarrow$ \ch{NH3 + HCN}. 
This peak in \ch{NH3} is at the same height as the dip in \ch{H2O} (look Figure \ref{fig:100x_carbon-water}), which reflects the use of \ch{OH} radicals. It also coincides with the net result of the domination of ammonia production over destruction at given height (see Figure \ref{fig:ammonia_rate}).  \ch{OH} radicals are also used in \ch{NO} production, which follows the same pathway as in \cite{Moses_2011}:\\

\hspace{-0.4cm} 3 x (\ch{H2O + $h\nu$} $\rightarrow$ \ch{OH + H})\\
\ch{NH3 + OH} $\rightarrow$ \ch{NH2 + H2O}\\
\ch{NH2 + OH} $\rightarrow$ \ch{NH + H2O}\\
\ch{NH + H} $\rightarrow$ \ch{N + H2}\\
\ch{N + OH} $\rightarrow$ \ch{NO + H}\\
$\frac{3}{2}$\Big(\ch{H + H} $\rightarrow$ \ch{H2 }\Big)\\
\textbf{Net:} \ch{NH3 + H2O + 3 $h\nu$} $\rightarrow$ \ch{NO + $\frac{5}{2}$ H2 }\\

In contrast to samples with \ch{N2}, HCN production at its peak occurs without OH radicals coming from \ch{H2O} and is the same as in \cite{Moses_2011}. Because the HCN peak is high in the atmosphere, its destruction occurs mainly through combination with H atoms and not OH radicals or less abundant O atoms like in \cite{Rimmer_2019}. High abundance of \ch{NH3} molecules allows HCN to be formed earlier than in samples with \ch{N2} (compare Figures \ref{fig:N2.2p_cold-trap} and \ref{fig:100x_nitrogen_sulfur}), however, its peak is at similar height as in \ch{N2} samples, because its main path of production follows from methyl radicals, which are produced at around $10^{-7}$ bar in both cases:\\

\hspace{-0.4cm} 2 x (\ch{NH3 + $h\nu$} $\rightarrow$ \ch{NH2 + H})\\
\ch{NH2 + NH2} $\rightarrow$ \ch{HNNH + H2}\\
\ch{HNNH + H} $\rightarrow$ \ch{N2H + H2}\\
\ch{N2H + $h\nu$} $\rightarrow$ \ch{N2 + H}\\
\ch{NH3 + $h\nu$} $\rightarrow$ \ch{NH* + H2}\\
\ch{NH* + M} $\rightarrow$ \ch{NH + M}\\
\ch{NH + H} $\rightarrow$ \ch{N + H2}\\
\ch{CH4 + H} $\rightarrow$ \ch{CH3 + H2}\\
\ch{CH3 + N} $\rightarrow$ \ch{H2CN + H}\\
\ch{H2CN + $h\nu$} $\rightarrow$ \ch{HCN + H}\\
\ch{H + H} $\rightarrow$ \ch{H2}\\
\textbf{Net:} \ch{3 NH3 + CH4 + 5 $h\nu$} $\rightarrow$ \ch{HCN + N2 + 6 H2}\\

The dissociation of water and methane, combined with the production of \ch{HCN}, leads to a sharp peak in cyanoacetylene (\ch{HC3N}) around $10^{-7}$ bar. This production pathway to \ch{HC3N} is facilitated through interstellar-like ammonia production from \ch{N2} dissociation in the upper atmosphere (see Figure \ref{fig:100x_nitrogen_sulfur}, where a peak in ammonia appears just below the peak of cyanoacetylene). Deeper in the atmosphere, ammonia is used up to form \ch{CH5N}, \ch{N2}, and HCN and does not contribute directly to \ch{HC3N} production. Only once the destruction of HCN happens, \ch{HC3N} can form more efficiently, hence we observe a sharper, higher situated peak in the atmosphere compared to \ch{N2} samples (see Figures \ref{fig:N2.2p_cold-trap} and \ref{fig:100x_nitrogen_sulfur}). Larger hydrocarbons (like \ch{C2H4}) begin to break apart around the same pressure and therefore further accelerate this process by providing extra \ch{C2H2} molecules:\\

\hspace{-0.4cm} \ch{C2H2 + CN} $\rightarrow$ \ch{HC3N + H}\\
\ch{HCN + $h\nu$} $\rightarrow$ \ch{CN + H}\\
\ch{H2CN + $h\nu$} $\rightarrow$ \ch{HCN + H}\\
\ch{N + CH3} $\rightarrow$ \ch{H2CN + H}\\
\ch{CH4 + OH} $\rightarrow$ \ch{CH3 + H2O}\\
\ch{CH4 + H} $\rightarrow$ \ch{CH + 2 H2}\\
\ch{CH4 + CH} $\rightarrow$ \ch{C2H4 + H}\\
\ch{C2H4 + $h\nu$} $\rightarrow$ \ch{C2H2 + H2}\\
\ch{H2O + $h\nu$} $\rightarrow$ \ch{OH + H}\\
\ch{N2 + $h\nu$} $\rightarrow$ \ch{N+ + N + e-}\\
\ch{N+ + H2 } $\rightarrow$ \ch{ NH+ + H}\\
\ch{NH+ + H2 } $\rightarrow$ \ch{ NH2+ + H}        \\
\ch{NH2+ + H2 } $\rightarrow$ \ch{ NH3+ + H}        \\
\ch{NH3+ + H2} $\rightarrow$ \ch{ NH4+ + H}      \\
\ch{NH4+ + e- } $\rightarrow$ \ch{ NH3 + H}        \\ 
5 x (\ch{H + H} $\rightarrow$ \ch{H2})\\
\textbf{Net:} \ch{3 CH4 + N2 + 5 $h\nu$} $\rightarrow$ \ch{HC3N + NH3 + 4 H2}\\

If ammonia is abundant (unlike in \ch{N2} samples) there is a very simple chemical pathway that leads to the production of \ch{CH5N}:\\

\hspace{-0.4cm} \ch{NH3 + $h\nu$ } $\rightarrow$ \ch{NH2 + H}\\
\ch{CH4 + H} $\rightarrow$ \ch{CH3 + H2}\\
\ch{CH3 + NH2} $\rightarrow$ \ch{CH5N}\\
\textbf{Net:} \ch{NH3 + CH4 + $h\nu$} $\rightarrow$ \ch{CH5N + H2}\\

The path is different from \cite{Rimmer_2019}, where \ch{CH5N} is formed directly through \ch{CH4 + NH + M} $\rightarrow$ \ch{CH5N + M}.\\

Large \ch{H2S} abundance at the surface leads to significant amounts of hazes due to sulphur species condensing out (Figure \ref{fig:100x_nitrogen_sulfur}). \ch{S2} condensate ('J' denotes condensed species) is the most abundant throughout the whole atmosphere and comes directly from \ch{H2S} and the H atoms from the destruction of \ch{NH3}:\\

\hspace{-0.4cm} \ch{NH3 + $h\nu$ } $\rightarrow$ \ch{ NH2 + H}\\
\ch{H2 + NH2 } $\rightarrow$ \ch{ NH3 + H}\\
3 x (\ch{H2S + H } $\rightarrow$ \ch{ HS + H2})\\
\ch{HS + HS } $\rightarrow$ \ch{ H2S + S}\\
\ch{S + HS } $\rightarrow$ \ch{ S2 + H}\\
\ch{S2 } $\rightarrow$ \ch{ JS2} \\ 
\textbf{Net:} \ch{ 2 H2S + $h\nu$} $\rightarrow$ \ch{ JS2 + 2 H2}\\

Sulfur hazes at the bottom of the atmosphere come from \ch{H2S} destruction and then the recombination of relevant products:\\

\hspace{-0.4cm} 4 x (\ch{H2S + $h\nu$ } $\rightarrow$ \ch{ HS + H})\\
8 x (\ch{H2S + H } $\rightarrow$ \ch{ HS + H2})\\
4 x (\ch{ HS + HS } $\rightarrow$ \ch{ H2S + S})\\
4 x (\ch{HS + S } $\rightarrow$ \ch{ S2 + H})\\
\ch{S2 + S2 } $\rightarrow$ \ch{ S4}\\
\ch{S2 + S4 } $\rightarrow$ \ch{ S6}\\
\ch{S2 + S6 } $\rightarrow$ \ch{ S8}\\
\textbf{Net:} \ch{ 8 H2S + 4 $h\nu$} $\rightarrow$ \ch{ S8 + 8 H2}\\ 

Even though sulfur and nitrogen species do not interact with each other, this significant production of sulfur haze in the atmosphere will likely impact the visibility of other compounds, hence it is important to investigate this result further.

\subsection{Sample with \ch{H2S}}\label{subsec:h2s}

\begin{figure}
    \captionsetup{width=\linewidth}
    \includegraphics[width=\linewidth]{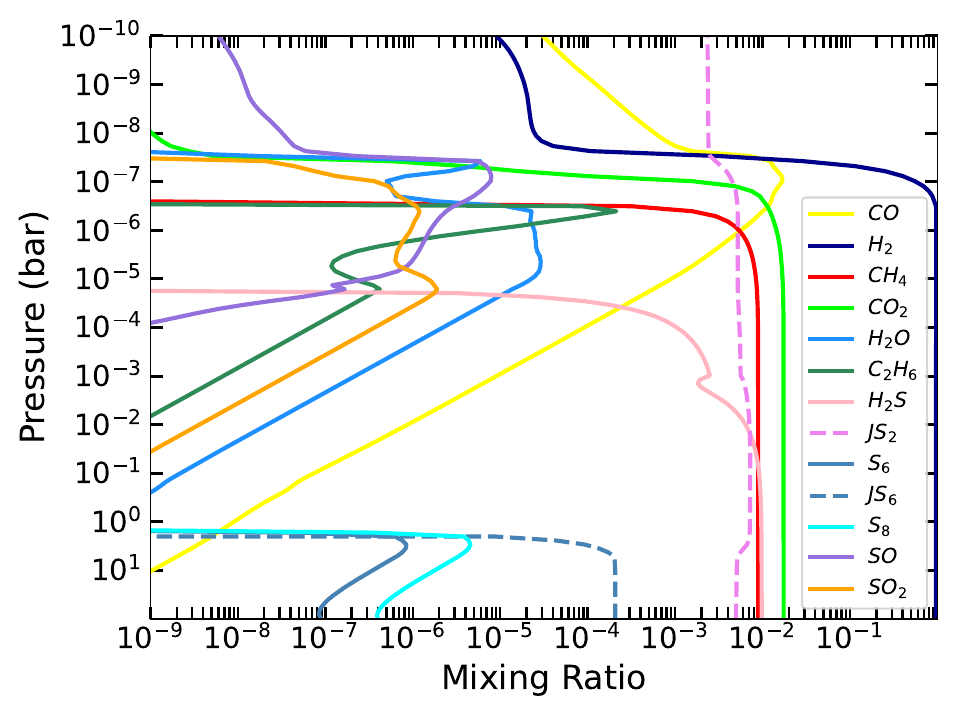}
    \caption{Height profile for $K_{zz}$ trap sample with 1\% of \ch{H2S} added. For clarity, there are only plotted molecules taking part in production of sulfur bearing species, however all the inputs and products from Figure \ref{fig:N2_0_cold trap} are also present. 'J' denotes condensed species.}
    \label{fig:H2S_2_ct}
\end{figure}

Sulfur hazes in the bottom of the atmosphere are produced by a simple conversion of \ch{H2S} molecules: 

\ch{2 H2S } $\rightarrow$ \ch{ S2 + 2 H2} - these \ch{S2} molecules then 'glue' together to produce larger compounds:  \ch{S2 + S2 } $\rightarrow$ \ch{ S4}, \ch{S2 + S4} $\rightarrow$ \ch{ S6} and finally \ch{S2 + S6 } $\rightarrow$ \ch{ S8}. They can then condense to produce the aforementioned haze (dashed lines on Figures \ref{fig:H2S_2_ct} and \ref{fig:100x_nitrogen_sulfur})

In the upper atmosphere, due to photodissociation of carbon dioxide, oxygen atoms are released and combine with S atoms to produce \ch{SO} and \ch{SO2}:\\

\hspace{-0.4cm} 2 x (\ch{CO2 + $h\nu$ } $\rightarrow$ \ch{ CO + O})\\
3 x (\ch{H2S + H } $\rightarrow$ \ch{ HS + H2})\\
2 x (\ch{HS + O } $\rightarrow$ \ch{ SO + H})\\
\ch{SO + SO } $\rightarrow$ \ch{ SO2 + S}\\
\ch{HS + S } $\rightarrow$ \ch{ S2 + H}\\
\ch{S2 } $\rightarrow$ \ch{ JS2}\\ 
\textbf{Net:} \ch{2 CO2 + 3 H2S + 2 $h\nu$} $\rightarrow$ \ch{ 2 CO + SO2 + JS2}\\

Overall, \ch{H2S} is the most sensitive specie to the presence of a $K_{zz}$ trap. Synthesis of oxygen and sulfur dominates the reactions of these species in the upper atmosphere. Condensed disulphur - \ch{JS2} - is the dominating sulfur compound in the atmosphere (see Figure \ref{fig:H2S_2_ct}). 

%% file: Discussion.tex
\section{Discussion}\label{sec:discussion}
The results show how the detection of nitrogen compounds could be used to determine the specific conditions on the exoplanet: dominance of \ch{NH3} indicates chemical equilibrium, while the lack of nitrogen species in the spectrum does not rule out the possibility of \ch{N2} being present. Which case better explains the retrieved data from K2-18b? Mini-Neptune (chemical equilibrium) interpretation by \cite{Wogan_2024} seems easier to accept due to its simplicity, however, it does not explain large \ch{CO2} amounts found in the atmosphere. 100x solar metallicity sample (Table \ref{tab:100x_solar}) predicts \ch{CO2} should be created in much lower quantities (Figure \ref{fig:100x_cold trap}) than detected. Additionally, there is no explanation in there for non-detection of ammonia, which could be only explained by a presence of a major sink for it (for example an ocean). Treating K2-18b as a hycean world would require high albedo and dry stratosphere to explain the non-detection of water on a supposedly water rich world. 100x solar metallicity sample carries \ch{H2O} into the stratosphere, however it could be due to rain out not being implemented in the model. \cite{Yang_2024} suggested that \ch{CO2}/\ch{CH4} ratio could be used to infer the amount of water in the planet and estimated that K2-18b should have an envelope with about 10\% of \ch{H2O}. The presence of both \ch{CH4} and \ch{CO2}, combined with the cool temperature profile used, might indicate there is a similarity between the chemistry of Titan and K2-18b. HCN formation and the recycling of hydrocarbons is the same as in \cite{Yung_1984}, but the formation of \ch{HC3N} in that study mainly comes from \ch{C2H + HCN} $\rightarrow$ \ch{HC3N + H}, while in our paper from \ch{C2H2 + CN}.

We found that \ch{C2H6} is produced in significant amounts in the atmosphere. It absorbs strongly in the infrared range and, if present in high quantities, we would expect it should be observable by JWST. \cite{Madhusudhan_2023} reported trace amounts of DMS. Since DMS break up results in release of \ch{CH3} radicals, it could increase the production of ethane. If \ch{C2H6} is ever detected with abundance larger than predicted in this work, it could be an argument for DMS presence in K2-18b's atmosphere.

According to \cite{schwieterman2017a} and \cite{Wogan_2024}, simultaneous presence of \ch{CO2} and \ch{CH4} in significant amounts, must imply biological source if the planet is a hycean world. These two gases coexisting could also mean abiotic water-rock interactions or C/O ratio very different from expected. Hycean world assumption limits this result to biological origin of either of these gases, since hundreds of km's of thick ice mantle does not allow for intense volcanism and rock tectonics (\cite{Kite_2018}). The simulations do not consider collisional induced absorption of \ch{H2} molecules (\cite{Innes_2023}) and since a colder temperature profile was used, the steam runaway greenhouse effect was not observed - the atmosphere stays optically thin in NIR Infrared (Figure \ref{fig:opt-depth}).

For samples with \ch{N2} and cold temperature profile, photochemistry does not produce enough N bearing species to approach detectable limit (around 1ppm). This is the case when considering both: the whole atmosphere and only detectable region (pressures between 10$^{-2}$ and 10$^{-5}$ bar - see Figure \ref{fig:N2_plots}). The distinction between what is produced and which products can be actually detected is very important, as most of nitrogen compounds are created in the upper atmosphere. This result has two implications: \\
1) We might not see nitrogen, even if it's present, although nitrogen species could potentially be detectable if the bulk nitrogen is in the form of \ch{NH3}. For species close to the detection limit, values found in this paper could be good predictions for future telescopes. \\
2) For hycean worlds, photochemical production of fixed nitrogen species like HCN is unlikely to produce surface concentrations relevant for prebiotic chemistry. Other sources of fixed nitrogen (e.g., impacts, lightning, volcanism) would be needed.

There are a few trends observed between all samples: most of them show peaks at around 1\% of added \ch{N2}. There are however subtle differences for changing $K_{zz}$ values. Within a certain range, results with different Eddy Diffusion profiles are comparable with each other (see Figure \ref{fig:N2_plots}). To illustrate on an example: \ch{HC3N} is highest at 2\% for $K_{zz} = 10^6$ and $K_{zz} = 10^5$ \unit{cm^2 s^{-1}}, but for the sample with cold trap, its peak is situated at 1\%. This change is due to larger production of \ch{HCN} (one of the \ch{HC3N} precursors) around $10^{-6}$ bar in the cold trap sample, but not in $K_{zz} = 10^6$ or $10^5$ \unit{cm^2 s^{-1}} samples.

Ammonia production is very interesting (see Section \ref{subsec:n2}) - the dominant cycle involves ions and is analogous to production of ammonia in the interstellar medium (\cite{Herbst}). It is unclear how likely it is for these reactions to actually occur in the atmosphere - the region in which they are prominent is purely in the extrapolated part of the pressure-temperature profile (see Section \ref{subsec:pT-profile}). In reality, there should be a thermosphere, where temperature would gradually increase, therefore, the conditions in the upper atmosphere would be very different from the interstellar medium and ammonia production would probably occur through a different path. Additionally, as interesting as the reaction is, the amount of ammonia produced is very tiny and is unlikely to be observationally constrained in the foreseeable future. \ch{NH3} can be used as a diagnostic for the atmospheric size, which has been addressed by \cite{Yu_2021}, \cite{Hu_2021}, \cite{Tsai_2021} - depletion of ammonia could occur in small atmospheres with liquid oceans.

\cite{Soni_2023} related the abundance of \ch{HCN} to the abundance of \ch{NH3} and the changing value of metallicity. A similar pathway was discovered here in 100x solar metallicity sample - \ch{HCN} is produced through photodissociation of ammonia and is directly proportional to the amount of \ch{NH3} present, however in other samples it is more related to the C\textbackslash O ratio, similarly to the findings of \cite{Rimmer_2019} and \cite{Mollier_2015}. \ch{HCN} then leads to production of \ch{HC3N}, which is more abundant than other compounds and hence more likely to be present in potentially detectable quantities. \cite{Rimmer_2021} also predicted there should be significant production of \ch{HC3N} in reducing, nitrogen rich atmospheres of super-Earths, which is comparable to the hydrogen rich sub-Neptune case with 10\% abundance of \ch{N2}. Although it is worth pointing out, the atmosphere they considered had a volcanic source of both HCN and \ch{C2H2}, which would explain the difference in \ch{HC3N} amounts between that paper and our current work. A spectral feature of \ch{HC3N} would appear around 4.5\unit{\mu m}, hence is accessible by JWST. Due to large abundance of hydrogen and carbon, \ch{HC3N} is a bad indicator of whether the atmosphere is in equilibrium or not, as it is produced in large amounts in samples with \ch{N2} and samples with 100x solar metallicity (see Figure \ref{fig:summary_plots}). If cyanoacetylene is detected in future observations, it could be compared with the results from this study. \cite{Blumenthal_2018} has shown that for cooler sub-Neptunes, disequilibrium chemistry can only be seen for planets with solar metallicity and that CO and \ch{CO2} should be both visible in the planet's spectrum. Although the presence of \ch{CO2} on K2-18b points to disequilibrium conditions, the non-detection of CO should be explained.

Finally, simulations with 100x solar metallicity (Table \ref{tab:100x_solar}) show that the presence of sulfur species has little to no effect on the abundance of nitrogen species, unlike haze from hydrocarbons, which participated in the production and recycling of nitrogen species. The production of haze with \ch{CH4} source was also considered in \cite{Kawashima_2019}. The significant production of sulfur hazes agrees well with the prediction of \cite{Madhusudhan_2021} and \cite{Piette_2020}. Detection of sulfur haze in the exoplanet could serve as a 'sulfur thermometer' - their formation is only possible in cold temperature profiles, hence their presence can limit the range of possible temperatures on the planet. If such hazes are present in large enough quantities, they could increase the planet's albedo enough to allow for liquid water to be present. In that case, it would be worth to consider the solubility of ammonia in the ocean and water clouds and whether it could explain its non-detection in \cite{Madhusudhan_2023}.

This study has not touched upon many interesting parameters that could impact the results, such as K2-18b's internal temperature, more complicated Eddy Diffusion profiles, different values of metallicity, hotter temperature-pressure profiles, inclusion of extra sinks for certain molecules at the surface or varying the pressure at which we would place the surface. 

The results suggest that the amounts of nitrogen species found in this study is not sufficient to approach detectable regimes. It is inconclusive whether it is due to simply low production in the atmosphere or dilution of such species in the ocean as suggested in \cite{Madhusudhan_2020} and \cite{cooke2024considerationsphotochemicalmodelingpossible}. Possible biosignature molecules for K2-18b were considered by \cite{Madhu_2025} and, in addition to DMS or diethyl sulfide, they found that methyl acrylonitrile could be a good candidate. Since nitriles are not formed from \ch{N2} in K2-18b's atmosphere, we would expect the abundance of \ch{C4H5N} to be much lower than HCN or \ch{HC3N}. It may be possible with an atmosphere rich in ammonia, but there is no good evidence of \ch{NH3} in the K2-18b's spectrum (\cite{hu2025waterrichinteriortemperatesubneptune}).

Recent measurements support our case for K2-18b possibly being a hycean planet with a dry stratosphere and hazy atmosphere (\cite{hu2025waterrichinteriortemperatesubneptune}), though this needs to be disentangled from the magma ocean scenario \citep{Shorttle2024}, which we do not explicitly explore here. The findings of \citet{hu2025waterrichinteriortemperatesubneptune} would point to nitrogen only coming from \ch{N2} due to non-detection of water or ammonia in the spectrum. However, further studies need to be done to explain the non-detection of CO or finding an alternate pathway to its destruction.

%% file: Conclusions.tex
\section{Conclusions}\label{sec:conclusions}

The question of whether K2-18b is habitable still remains open. The lack of knowledge about the exact temperature profile, surface conditions, and stellar flux data limits the precision to which we can perform the atmospheric retrieval accurately. Apart from atmospheric model, there also remains a problem of criteria for planet's habitability. One would need to consider which biomarkers are reliable and what environmental conditions are needed for life to form and survive long enough, so that we can measure it. 

The field of atmospheric characterization of exoplanets is growing rapidly and new measurements could change the interpretation of our results. Regardless, they provide a good constraint on some of the planetary parameters and elemental abundances.
This study has shown that:
\begin{enumerate}
\item \ch{NH3} is much more effective in producing nitrogen species than \ch{N2}. Possible future detection of \ch{HCN}, \ch{NH3} or \ch{CH5N} would mean that ammonia is the source of nitrogen in the planet. However \ch{HC3N} and \ch{NO} are predicted to have similar abundance in both \ch{NH3} and \ch{N2} cases, therefore they cannot be used to distinguish between the sources of nitrogen in the atmosphere. 

\item Simulations above show that photochemistry alone is insufficient in models with \ch{N2}, to produce enough \ch{NH3}, \ch{HCN} or \ch{HC3N} to approach detectable quantities – if these species appear in the future data, there must be an additional source of these molecules, the planet must have deeper, hotter surface to allow for thermochemical recycling of such species or their source gas is ammonia.

\item Chemical equilibrium at the surface (mini-Neptune interpretation) does not explain over 1\% \ch{CO2} abundance measured by JWST - this result could only be explained by subsolar C/O ratio or an additional \ch{CO2} flux. \ch{CO} could potentially be detectable in small amounts in range probed by JWST – non-detection could mean there is an additional sink for this molecule, or it was simply overpowered by dominant methane and carbon dioxide spectrum.

\item Significant organic and sulfur hazes could be present in the atmosphere. It will be worth investigating how much effect they have on planet's albedo and overall dynamics of the atmosphere. These hazes could obstruct the view for other molecules and the detection of their presence could be a good proxy for a cold temperature profile. 

\item The appearance of ethane in every sample suggests a similar study could be done by varying \ch{CH4} and \ch{CO2} inputs to test the overall production of \ch{C2H6} and use the results as a proxy for DMS presence - if ethane would be measured in larger quantities than expected, one could test the hypothesis by \cite{Domagal_2011} that it would come from \ch{CH3} radicals released in DMS dissociation.
\end{enumerate}

%% file: Acknowledgments.tex
P.~B.~R. received funding for this work from the Leverhulme Centre for Life in the Universe Joint Collaborations Research Project Grant G112026, Project KKZA/237.

Authors would like to thank Prof. N. Madhusudhan for providing the temperature-pressure profile for hycean K2-18b, as well as advice received both during the course of the project and writing of the manuscript. We would like to thank anonymous reviewers for valuable comments and help in improvement to the manuscript.

%% file: Appendix_1.tex
\section{Reference Sample}\label{Appendix_1}
\begin{figure*}
     \centering
     \begin{subfigure}[b]{0.48\linewidth}
    \includegraphics[width=\textwidth]{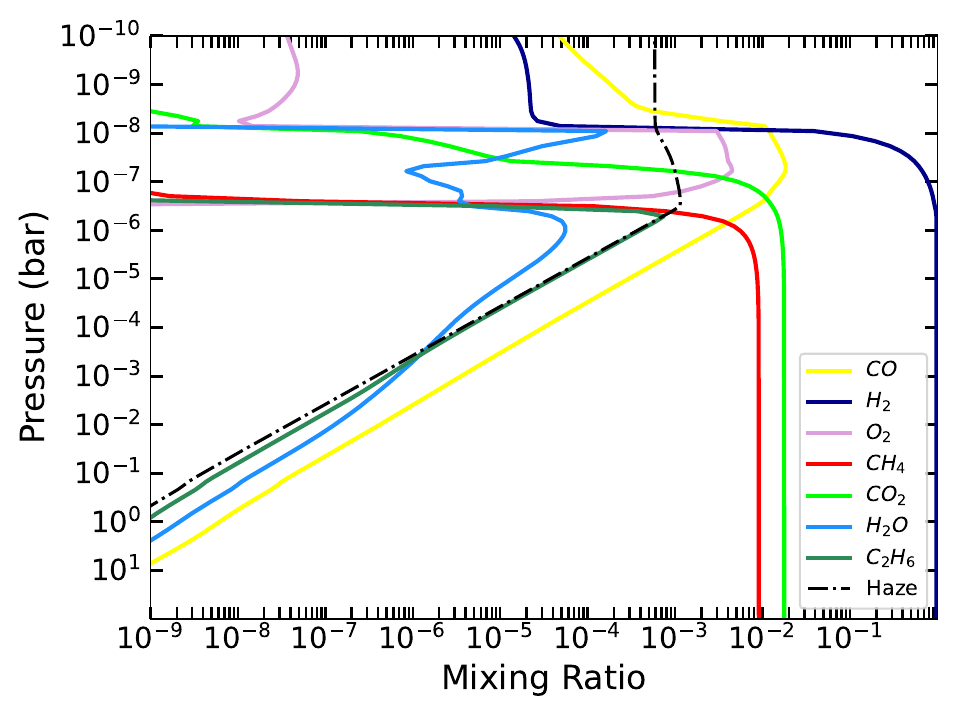}
    \caption{\textbf{GJ-436} - height profile for key molecules.}
    \label{fig:N2_0_gj436_sub}
     \end{subfigure}
     \hfill
     \begin{subfigure}[b]{0.48\linewidth}
    \includegraphics[width=\textwidth]{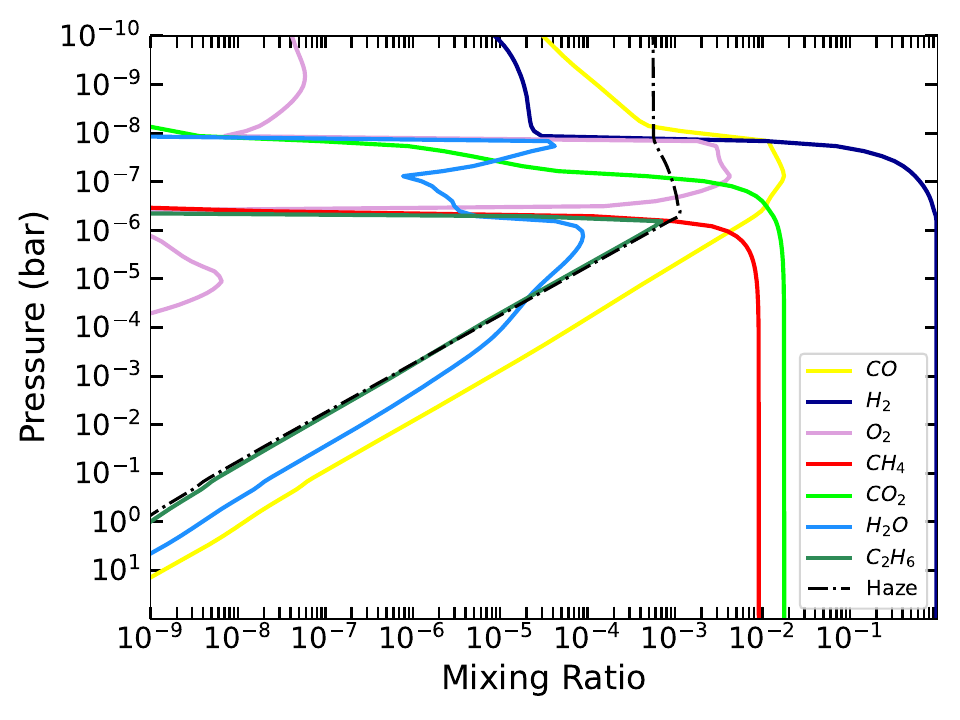}
    \caption{\textbf{GJ-176} - height profile for key molecules.}
    \label{fig:N2_0_Kzz6_sub}
    \end{subfigure}
    \vfill
    \begin{subfigure}{0.48\linewidth}
    \includegraphics[width=\textwidth]{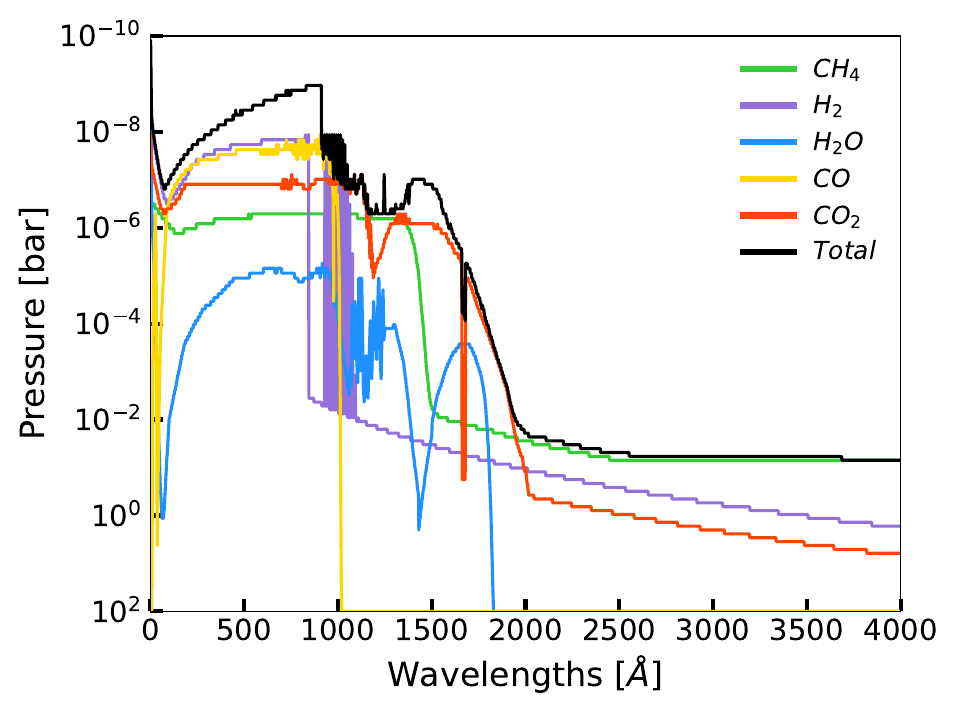}
    \caption{\textbf{GJ-436} - $\tau(\lambda) = 1$ for different wavelengths.}
    \label{fig:opt-depth_N2_0_gj436}
     \end{subfigure}
     \hfill
     \begin{subfigure}{0.48\linewidth}
    \includegraphics[width=\textwidth]{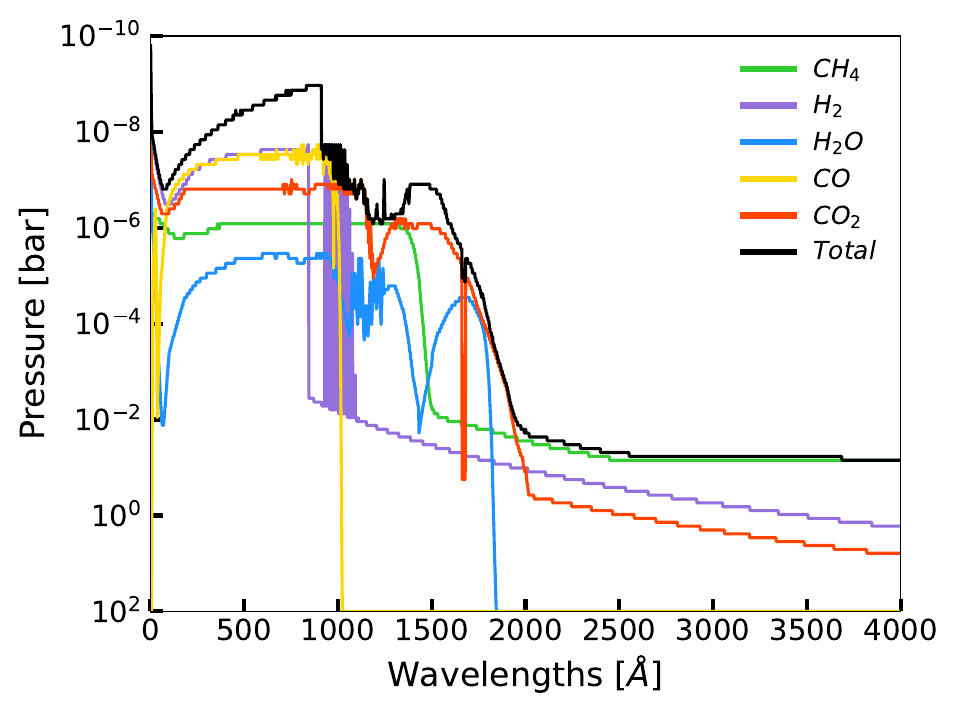}
    \caption{\textbf{GJ-176} - $\tau(\lambda) = 1$ for different wavelengths.}
    \label{fig:opt-depth_N2_0_Kzz6}
     \end{subfigure}
        \caption{The comparison of reference sample with $K_{zz} = 10^6$ \unit{cm^2 s^{-1}} between GJ-176 and GJ-436. The difference between stars is not significant and only visible in the upper atmosphere, where UV flux has the biggest impact. It shows that for the purpose of our calculations, the choice of star is arbitrary, as long as it is the same stellar type.}
        \label{fig:ref_stars}
\end{figure*}

\begin{figure*}
     \centering
     \begin{subfigure}[b]{0.48\linewidth}
    \includegraphics[width=\linewidth]{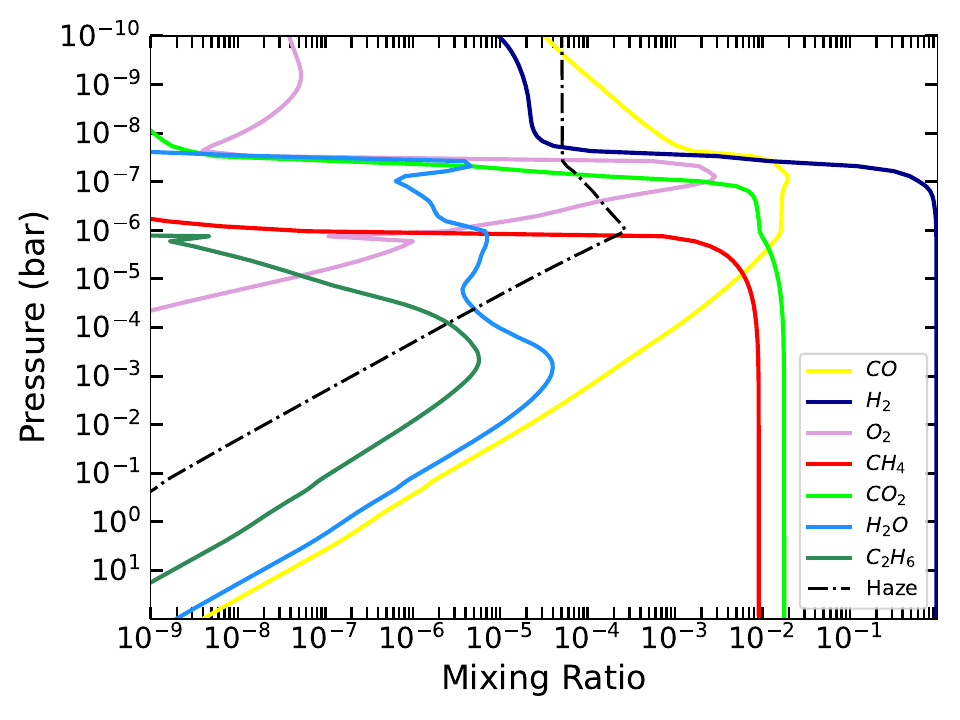}
    \caption{\textbf{$K_{zz}=10^4$ \unit{cm^2 s^{-1}}}}
    \label{fig:N2_0_Kzz4}
     \end{subfigure}
     \begin{subfigure}[b]{0.48\linewidth}
    \includegraphics[width=\linewidth]{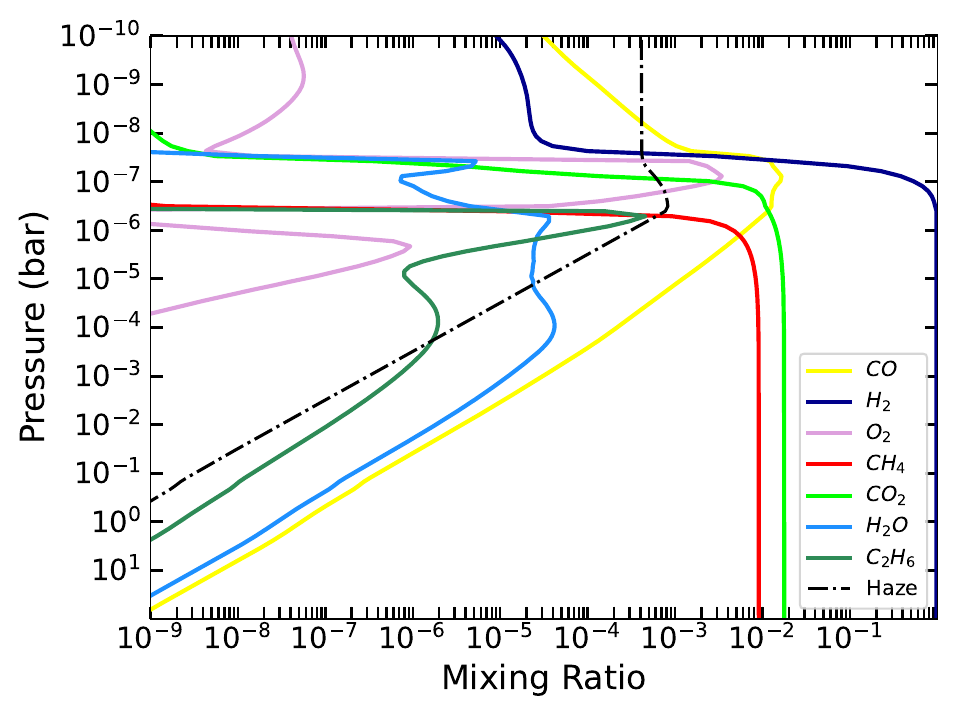}
    \caption{\textbf{$K_{zz}=10^5$ \unit{cm^2 s^{-1}}}}
    \label{fig:N2_0_Kzz5}
     \end{subfigure}
      \begin{subfigure}[b]{0.48\linewidth}
    \includegraphics[width=\linewidth]{k2-18b_reference_Kzz6.pdf}
    \caption{\textbf{$K_{zz}=10^6$ \unit{cm^2 s^{-1}}}}
    \label{fig:N2_0_Kzz6}
     \end{subfigure}
     \begin{subfigure}[b]{0.48\linewidth}
    \includegraphics[width=\linewidth]{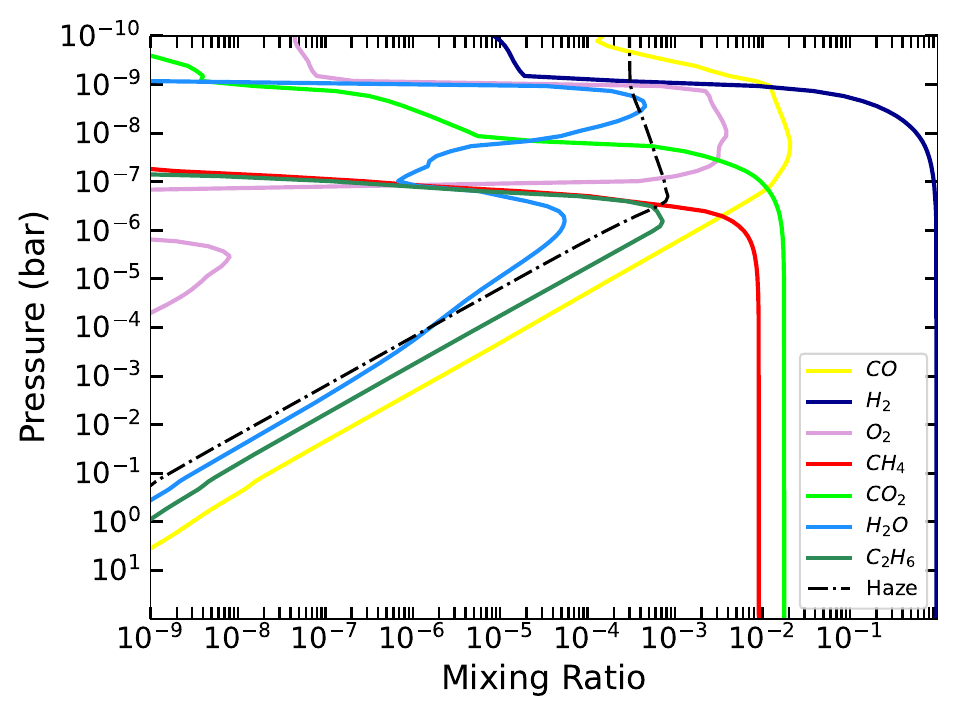}
    \caption{\textbf{$K_{zz}=10^7$ \unit{cm^2 s^{-1}}}}
    \label{fig:N2_0_Kzz7}
     \end{subfigure}
     \begin{subfigure}[b]{0.48\linewidth}
    \includegraphics[width=\linewidth]{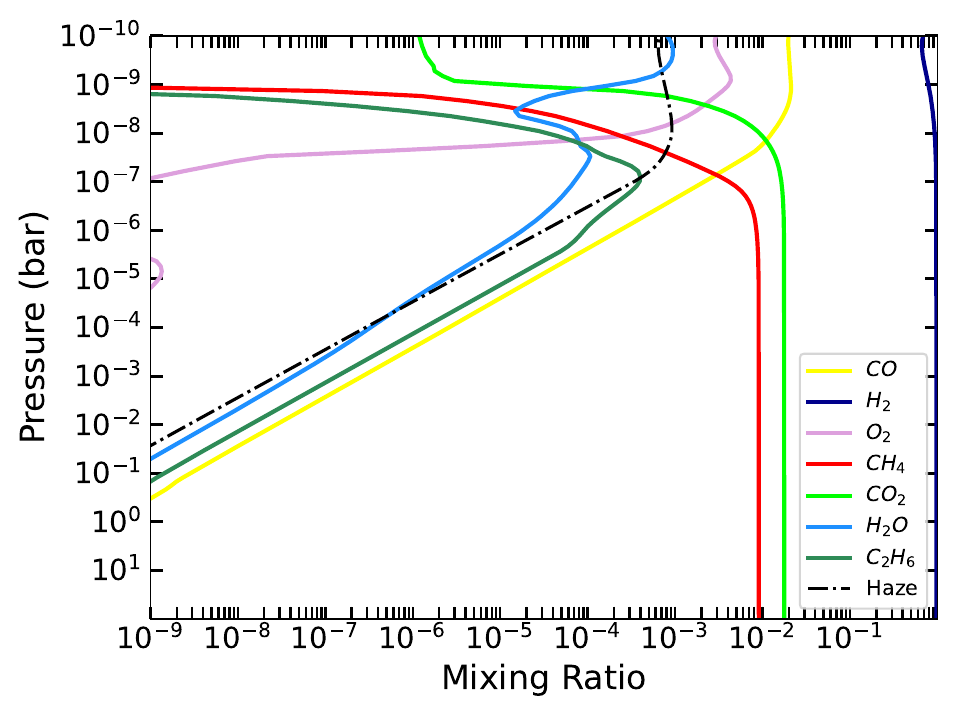}
    \caption{\textbf{$K_{zz}=10^8$ \unit{cm^2 s^{-1}}}}
    \label{fig:N2_0_Kzz8}
     \end{subfigure}
     \begin{subfigure}[b]{0.48\linewidth}
    \includegraphics[width=\linewidth]{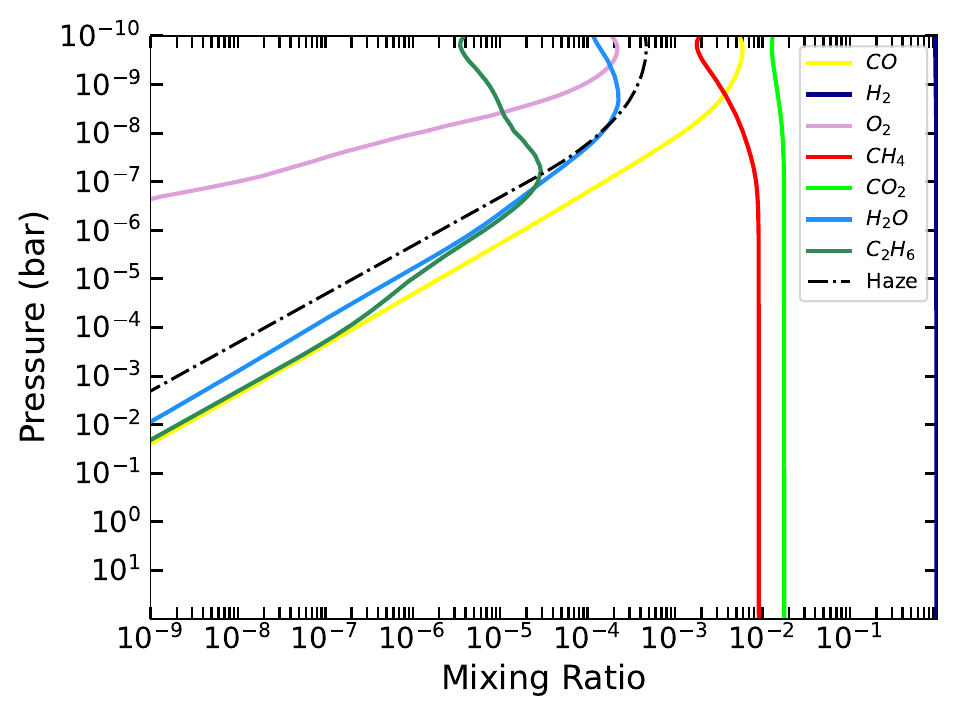}
    \caption{\textbf{$K_{zz}=10^9$ \unit{cm^2 s^{-1}}}}
    \label{fig:N2_0_Kzz9}
     \end{subfigure}
        \caption{The comparison of reference samples with varying $K_{zz}$ profiles. Reference sample with cold trap is shown on Figure \ref{fig:N2_0_cold trap}. Lower $K_{zz}$ values allow slower chemical reactions to occur in the deeper part of the atmosphere, while for higher $K_{zz}$ values, reactions will happen at the top of the atmosphere - the parcel of air is picked up closer to the surface and carried upwards, until it reaches more UV active region in which the molecules start to break apart.}
        \label{fig:reference_Kzz}
\end{figure*}

Different conditions were tested for the reference sample to check how sensitive are the key molecules to the variability in stellar spectrum or Eddy Diffusion. Figure \ref{fig:ref_stars} shows the comparison between the stars: GJ-176 and GJ-436. Figure \ref{fig:reference_Kzz} shows the impact of $K_{zz}$ profile on the height profiles of molecules from the reference sample.

\begin{figure*}
     \centering
     \begin{subfigure}[b]{0.48\linewidth}
    \includegraphics[width=\textwidth]{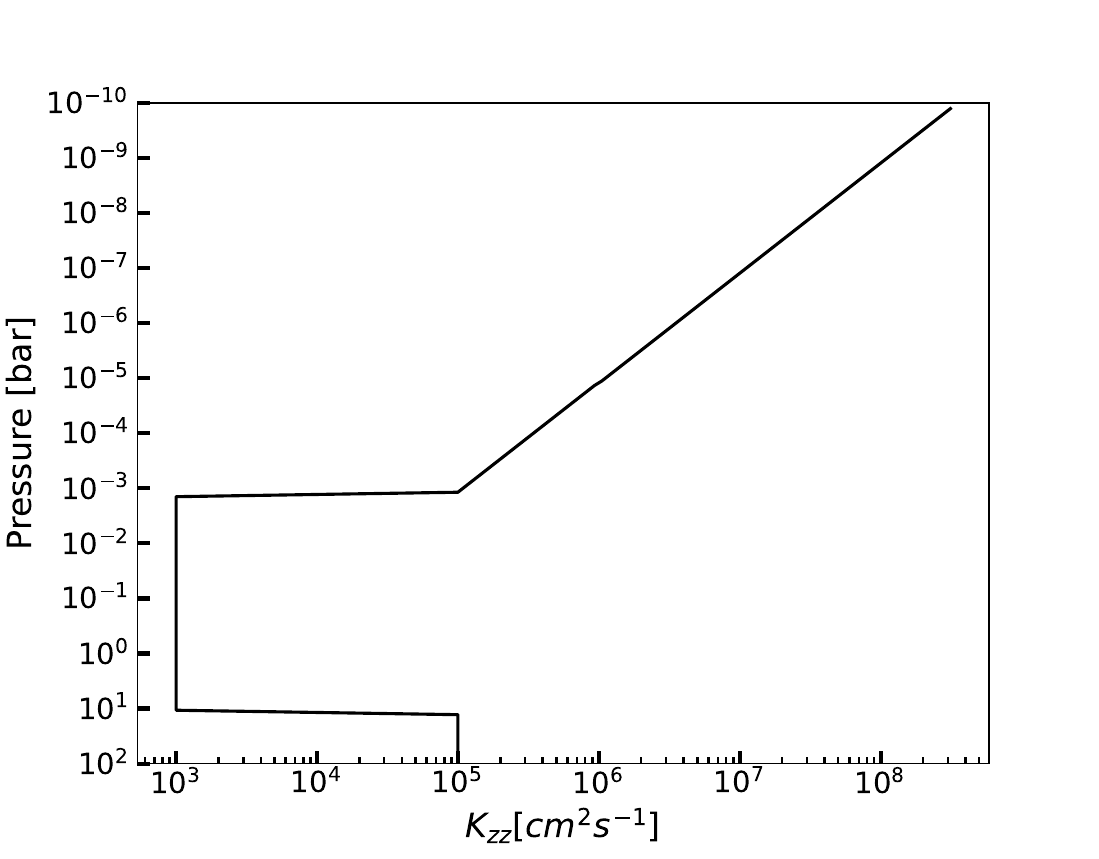}
    \caption{Another $K_{zz}$ profile tested with $K_{zz}$ following in the stratosphere $n^{-\frac{1}{2}}$, $n$ being the gas number density.}
    \label{fig:Kzz_sqrt}
     \end{subfigure}
     \hfill
     \begin{subfigure}[b]{0.48\linewidth}
    \includegraphics[width=\textwidth]{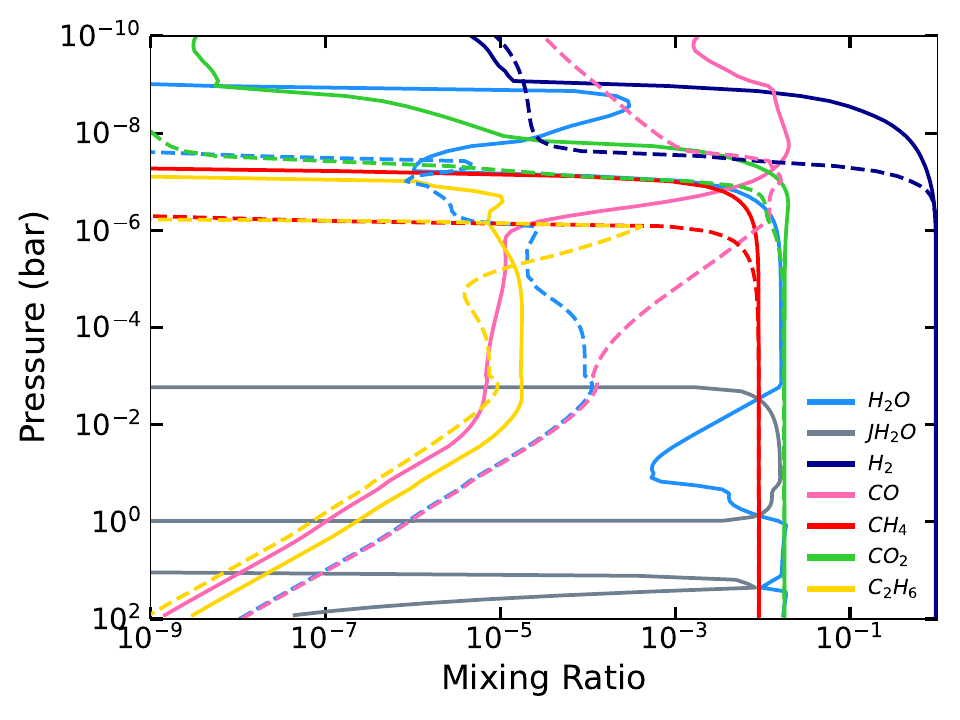}
    \caption{Comparison between previous $K_{zz}$ profile - solid lines ($K_{zz} = 10^5$ \unit{cm^2 s^{-1}} with ‘$K_{zz}$ trap’) and $K_{zz}$ increasing with $n^{-\frac{1}{2}}$ - dashed lines.}
    \label{fig:Kzz_comparison}
    \end{subfigure}
        \caption{Testing more complex model of Eddy Diffusion - $K_{zz}$ increasing with $n^{-\frac{1}{2}}$ ($n$ being the gas number density). Deviations between these cases from Figures \ref{fig:Kzz} and \ref{fig:Kzz_sqrt} are prominent only in the upper part of the atmosphere, above the detection range.}
        \label{fig:ref_Kzz}
\end{figure*}

We have also tested a model with $K_{zz}$ profile with an increasing Eddy Diffusion coefficient (see Figure \ref{fig:Kzz_sqrt}). We have found that there is little difference in the detectable region of the atmosphere from our previous $K_{zz}$ profile (see Figure \ref{fig:Kzz_comparison}).

%% file: Appendix_5.tex
\section{Comparison of dry and wet samples}\label{Appendix_5}
\begin{figure}
    \centering
    \includegraphics[width=0.75\linewidth]{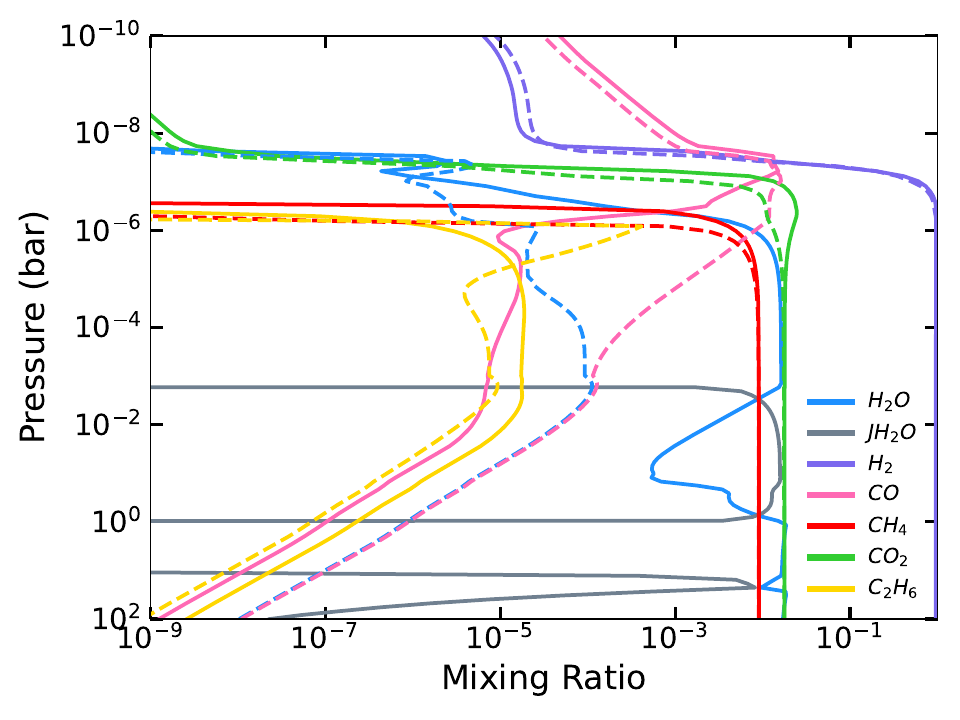}
    \caption{Reference sample with water added (solid lines) and without water (dashed lines). \ce{JH2O} marks water in the condensed phase.}
    \label{fig:ref_sample_water_comparison}
\end{figure}

\begin{figure}
    \centering
    \includegraphics[width=0.75\linewidth]{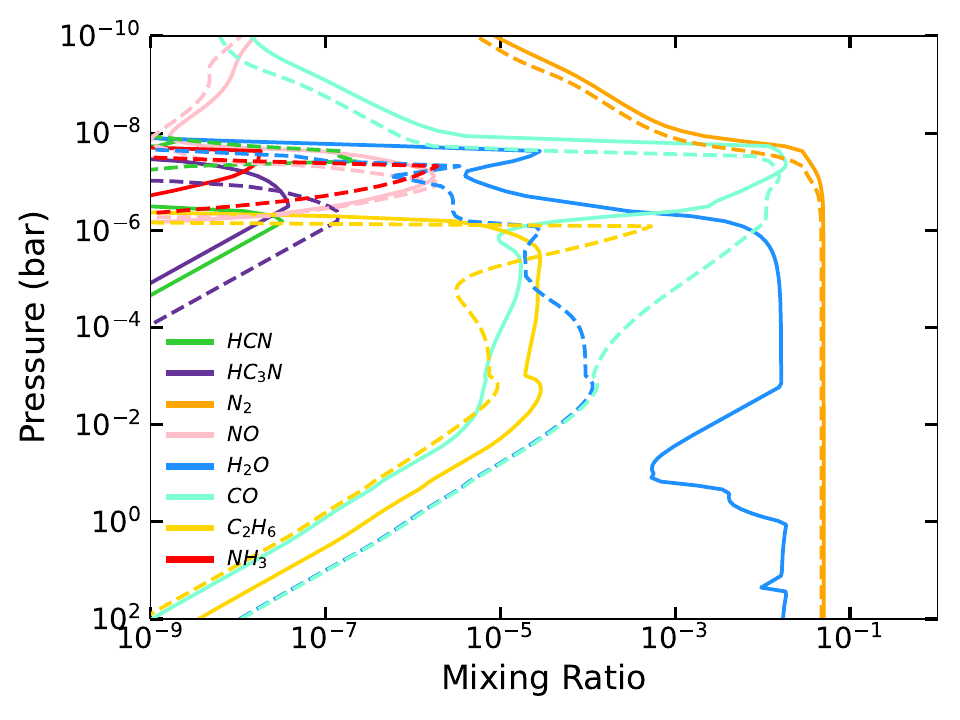}
    \caption{Sample with 5$\%$ of added \ce{N2} and water at saturation pressure (solid lines) and without water (dashed lines).}
    \label{fig:N2_sample_water_comparison}
\end{figure}

The presence of water had minimal impact on the reactions of nitrogen compounds in the atmosphere for samples with \ce{N2}. As can be seen on Figures \ref{fig:ref_sample_water_comparison} and \ref{fig:N2_sample_water_comparison}, the addition of water mostly affected the shape of \ce{C2H6}, CO and HCN curves. Nevertheless, the peak values of all compounds were larger in the dry case scenario.

The production of ethane in the wet sample is independent of the photolysis of \ce{CO2} to get to the OH radicals. Hence, the amount of \ce{C2H6} in the deeper atmosphere is larger, when water is present. See section \ref{subsec: reference} and Figure \ref{fig:ref_sample_water_comparison} for comparison. \\
 
\hspace{-0.4cm} \ce{H2O + $h\nu$} $\rightarrow$ \ce{OH + H} \\
\ce{H2 + OH} $\rightarrow$ \ce{H2O + H} \\
2 x \Big(\ce{CH4 + H} $\rightarrow$ \ce{CH3 + H2}\Big)\\
\ce{CH3 + CH3} $\rightarrow$ \ce{C2H6}\\
\textbf{Net:} \ce{2 CH4 + $h\nu$} $\rightarrow$ \ce{C2H6 + H2}.\\

CO production also starts with the break up of \ce{CO2} by photons, but the whole production cycle goes through a combination of methane and OH radicals coming from water:\\

\hspace{-0.4cm} \ce{CO2 + $h\nu$} $\rightarrow$ \ce{CO + O} \\
\ce{H2O + $h\nu$} $\rightarrow$ \ce{OH + H}\\
\ce{CH4 + OH} $\rightarrow$ \ce{CH3 + H2O} \\
\ce{CH3 + O} $\rightarrow$ \ce{CH2O + H} \\
\ce{CH2O + H} $\rightarrow$ \ce{CHO + H2}\\
\ce{CHO + H} $\rightarrow$ \ce{CO + H2}\\
\textbf{Net:} \ce{CO2 + CH4 + 2 $h\nu$} $\rightarrow$ \ce{2 CO + 2 H2}.\\

Dry atmospheres exhibit ion-neutral chemistry and lead to the formation of \ce{HCN} on a path not very dependent on methane. It is a surprise that the chemistry looks positively interstellar. The reaction for ammonia in Section \ref{subsec:N2 vs NH3}, for example, is a mirror of the path for interstellar ammonia. At the same time, analogous to the chemistry of a photodissociation region in the interstellar medium, carbon dioxide breaks apart as \\

\hspace{-0.4cm} \ch{CO + $h\nu$} $\rightarrow$ \ch{C+ + O + e-}.\\
And then:\\
\hspace{-0.4cm} \ch{C+ + NH3} $\rightarrow$ \ch{CH2N+ + H} \\ 
\ch{CH2N+ + e-} $\rightarrow$ \ch{HCN + H}.\\

This path is utterly disrupted by the hydroxyl radicals and more freely available oxygen when water is abundant, especially if the water is not effectively trapped. However, a more robust formation mechanism for \ce{HCN} persists, and this other formation path persists when the atmosphere is dry and wet, and leads to similar results for each. Namely, a path the second half of which matches the classic path from \citet{Pearce2020}:\\

\hspace{-0.4cm} \ch{CH4 + $h\nu$} $\rightarrow$ \ch{CH2* + H2}, \\
\ch{CH2* + H2} $\rightarrow$ \ch{CH3 + H}, \\
\ch{CH3 + N} $\rightarrow$ \ch{H2CN + H}, \\
\ch{H2CN + M} $\rightarrow$ \ch{HCN + H + M},\\
$\frac{1}{2}$\big(\ch{N2 + $h\nu$} $\rightarrow$ 2\ch{N}\big)\\
$\frac{3}{2}$\big(2\ch{H + M} $\rightarrow$ \ch{H2 + M}\big)\\
\textbf{Net:} \; \ch{CH4 + $\frac{1}{2}$ N2 + $h\nu$} $\rightarrow$ \ch{HCN + $\frac{3}{2}$ H2}.\\

If this was all that occurred, then the results between the wet and dry cases would be nearly identical. But they differ by an order of magnitude for the nitriles. This is for two reasons. First, because of hydroxide radicals, the atomic nitrogen is more rapidly destroyed at critical atmospheric heights where it would otherwise participate in the formation of cyanide. Namely:\\

\hspace{-0.4cm} \ch{N + OH} $\rightarrow$ \ch{NO + H}, \\
\ch{NO + N} $\rightarrow$ \ch{N2 + O}.\\

\begin{figure}
    \centering
    \includegraphics[width=0.75\linewidth]{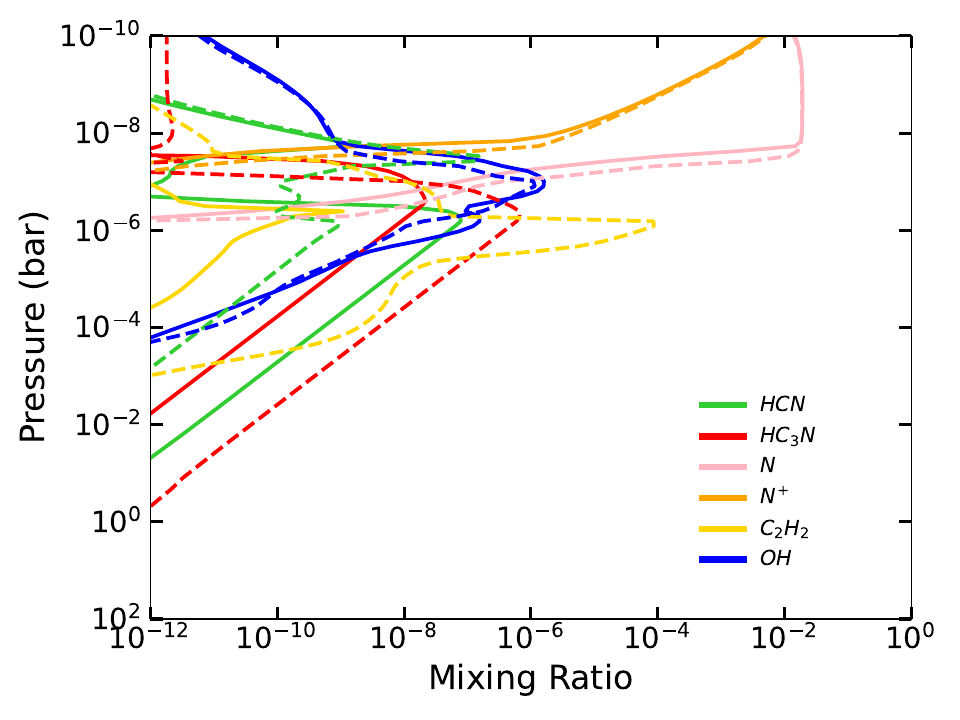}
    \caption{Sample with 2$\%$ of added \ce{N2} and water at saturation pressure (solid lines) and without water (dashed lines). Shown are molecules relevant for the production and destruction of HCN. The order of magnitude difference in the amount of \ce{HC3N} produced in the dry samples is responsible for the destruction (and hence lower amount) of surviving HCN, compared with wet samples. For full comparison of production pathways, see Section \ref{sec:results}.}
    \label{fig:N2_HCN_comparison}
\end{figure}

Each hydroxide radical removes two \ce{N}'s. In addition, hydroxide radicals deplete acetylene by several orders of magnitude, meaning that cyanoacetylene, which achieves abundances of parts per billion in the dry case, is on the order of parts per trillion in the wet case. When accounting for both cyanide and cyanoacetylene, our major nitriles, the difference between the dry and wet cases are about one order of magnitude more nitriles in the dry case than in the wet case. See the Tables \ref{tab:ratio_dry_wet1} \& \ref{tab:ratio_dry_wet2} and Figures \ref{fig:N2_sample_water_comparison} \& \ref{fig:N2_HCN_comparison}. Since nitriles were not abundant enough to be observed even in the more favourable dry case, our predictions remain the same for wet and dry atmospheres: no observable nitriles.

\begin{table}
    \centering
    \begin{tabular}{||cccc||}
    \hline
    \multicolumn{4}{||c||}{\textbf{Reference sample with water}} \\
    \hline\hline
       \textbf{ \ce{H2}} & \textbf{\ce{CO2}} &  \textbf{\ce{CH4}} & \textbf{\ce{H2O}}\\
        \hline
          95.676\% & 1.780\% & 0.912 \% & 1.632\%\\
    \hline
    \end{tabular}
    \caption{\textbf{Reference sample with water at saturation pressure} - \ce{CO2}, \ce{CH4} and \ce{H2O} amounts do not change for all nitrogen samples with water, but \ce{H2} mixing ratio is reduced by the amount of added \ce{N2}. Values for \ce{CO2} and \ce{CH4} are taken from Table 2 (No Offset values) in \protect\cite{Madhusudhan_2023}. Saturation vapour pressure of water is calculated based on data from \protect\cite{Bridgeman_1964}.}
    \label{tab:reference_sample_h2o}
\end{table}

\begin{table}
    \centering
    \begin{tabular}{||c|ccccc||}
    \hline
       \textbf{ added \ce{N2}} & \textbf{\ce{NH3}} &  \textbf{\ce{HCN}} & \textbf{\ce{HC3N}} & \textbf{\ce{NO}} & \textbf{\ce{CH5N}}\\
        \hline
          10\% & 102 & 25.1 & 1.31 & 1.21 & 0.329 \\
          5\% & 105 & 7.18 & 3.71 & 1.02 & 5.75 \\
          2\% & 67.0 & 1.92 & 30.0 & 0.976 & 31.9 \\
          1\% & 43.6 & 1.05 & 16.4 & 0.935 & 38.1 \\
          0.1\% & 6.24 & 0.289 & 3.94 & 1.05 & 27.0 \\
          100ppm & 1.61 & 0.089 & 10.0 & 1.26 & 21.4 \\
          10ppm & 1.67 & 0.024 & 11.4 & 1.44 & 30.9 \\
          
    \hline
    \end{tabular}
    \caption{Ratio of peak amount of nitrogen specie in dry samples to peak amount of nitrogen specie in wet samples rounded to 3 significant figures. Overall tendency is that the amount of nitrogen is higher in dry samples compared to wet samples. However, it can vary between species - HCN in samples with lower amounts of added  \ce{N2} behaves differently.}
    \label{tab:ratio_dry_wet1}
\end{table}

\begin{table}
    \centering
    \begin{tabular}{||c|ccccc||}
    \hline
       \textbf{ added \ce{N2}} & \textbf{\ce{NH3}} &  \textbf{\ce{HCN}} & \textbf{\ce{HC3N}} & \textbf{\ce{NO}} & \textbf{\ce{CH5N}}\\
        \hline
          10\% & 116 & 0.0444 & 2.59 & 1.19 & 0.632 \\
          5\% & 85.7 & 0.0161 & 7.08 & 0.999 & 10.7 \\
          2\% & 52.9 & 0.0224 & 63.3 & 0.898 & 49.4 \\
          1\% & 31.0 & 0.0286 & 35.6 & 0.857 & 58.7 \\
          0.1\% & 4.55 & 0.0134 & 9.44 & 0.866 & 42.2 \\
          100ppm & 1.63 & 0.00387 & 24.2 & 0.935 & 34.8 \\
          10ppm & 1.64 & 0.00105 & 27.5 & 0.976 & 46.1 \\
          
    \hline
    \end{tabular}
    \caption{Ratio of whole atmospheric column amount of nitrogen specie in dry samples to column amount of nitrogen specie in wet samples rounded to 3 significant figures. Overall tendency is that the amount of nitrogen is higher in dry samples compared to wet samples. However, it can vary between species - HCN in samples with lower amounts of added  \ce{N2} behaves differently.}
    \label{tab:ratio_dry_wet2}
\end{table}

%% file: Appendix_4.tex
\section{Supplementary figures and tables}\label{Appendix_4}
\begin{table}
    \centering
    \begin{tabular}{||ccccccc||}
    \hline
    \multicolumn{7}{||c||}{\textbf{Haze}} \\
    \hline
       \ch{C3H3} & \ch{C3H4} & \ch{C3H5} & \ch{C4H} & \ch{C4H2} & \ch{C4H3} & \ch{C4H4}\\
    \hline
    \end{tabular}
    \caption{\textbf{Haze} - hydrocarbons plotted together as organic haze.}
    \label{tab:Haze}
\end{table}
\begin{figure}
    \centering
    \includegraphics[width=0.75\linewidth]{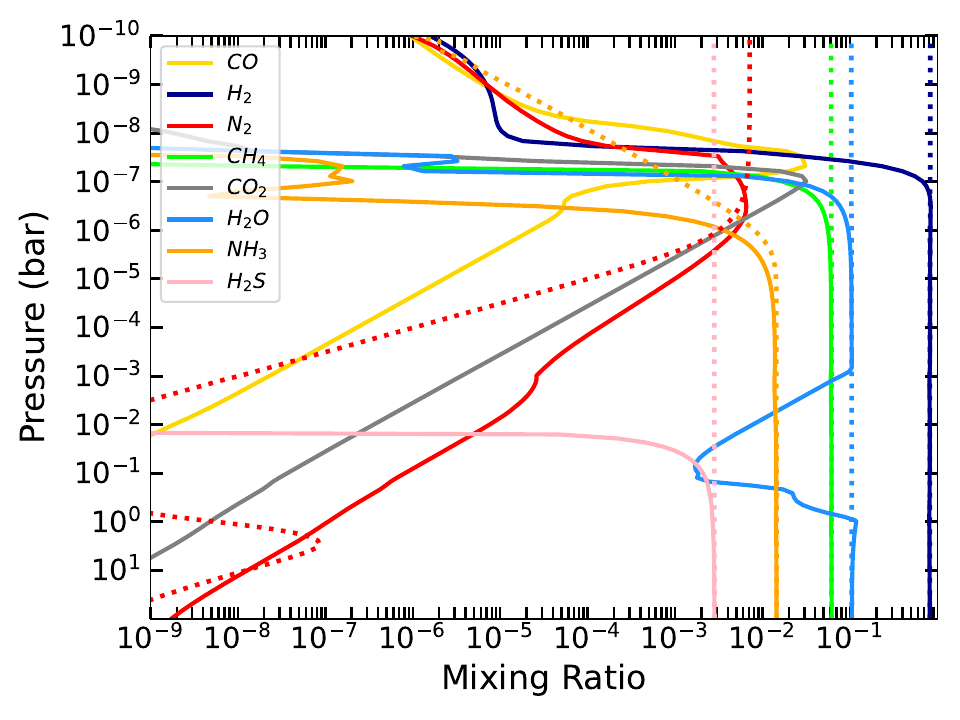}
    \caption{Height profile for cold trap sample with 100x solar metallicity. Dotted curves show chemical equilibrium curves calculated using \href{https://github.com/exoclime/FastChem}{FastChem} \citep{2022MNRAS.517.4070S} and solid curves show ARGO output. This figure shows how photochemistry impacts the abundance of species in the top part of the atmosphere. The lack of visible equilibrium curves for \ch{CO2} and CO is due to the fact that they do not reach the mixing ratio of 1ppb. This disequilibrium between \ch{CH4} and \ch{CO2} could indicate a different C/O ratio (other than solar) or a biosignature on a hycean planet.} 
    \label{fig:100x_cold trap}
\end{figure}
\begin{figure*}
    \centering
    \begin{subfigure}[b]{0.48\linewidth}
        \includegraphics[width=\linewidth]{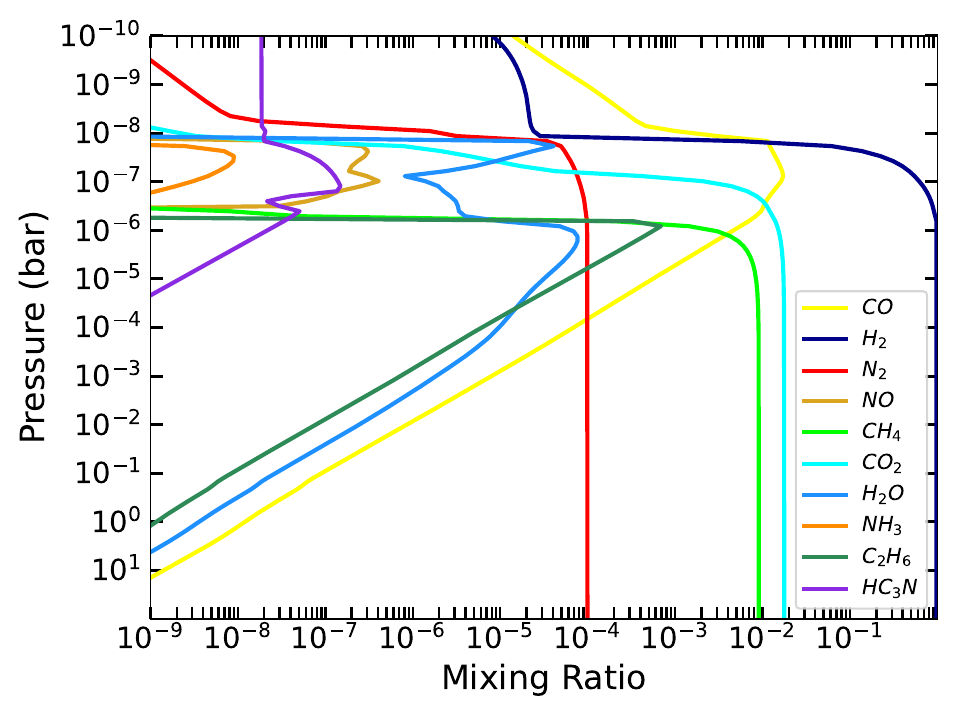}
        \caption{\textbf{\ch{N2}} - $10^{-4}$}
        \label{fig:N2_4_Kzz6}
    \end{subfigure}
    \hfill
    \begin{subfigure}[b]{0.48\linewidth}
        \includegraphics[width=\linewidth]{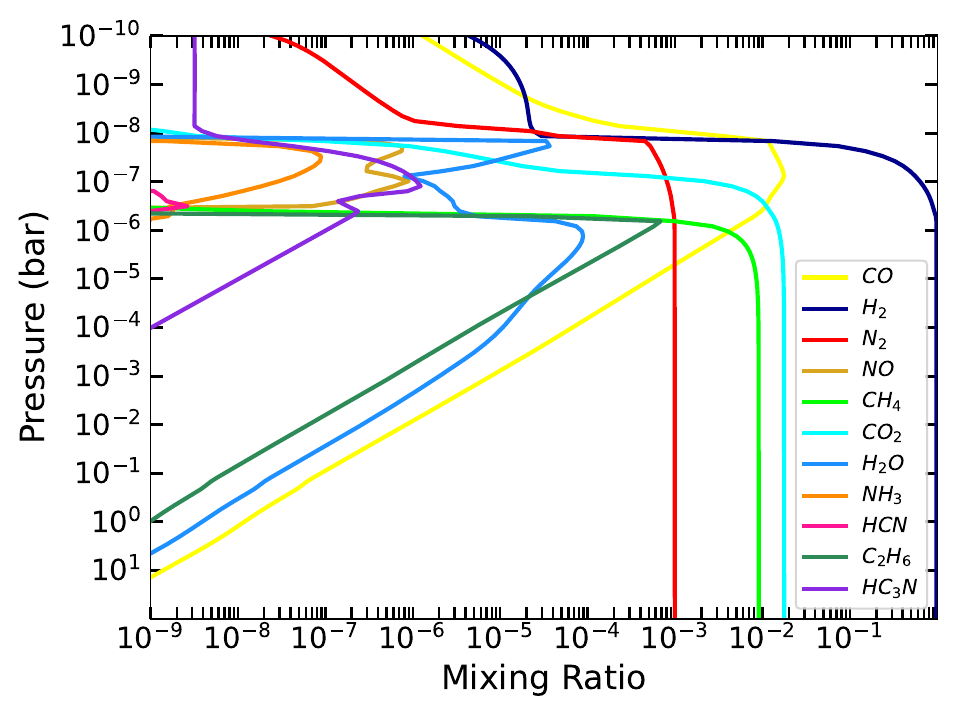}
        \caption{\textbf{\ch{N2}} - $10^{-3}$}
        \label{fig:N2_3_Kzz6}
    \end{subfigure}

    \vspace{1em} 

    \begin{subfigure}[b]{0.48\linewidth}
        \includegraphics[width=\linewidth]{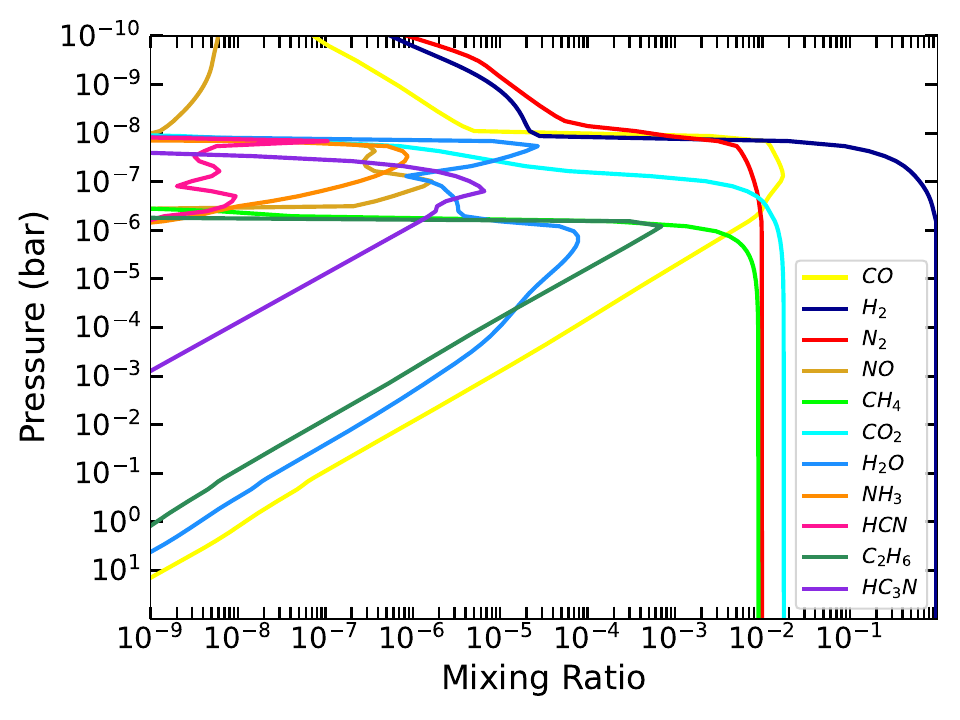}
        \caption{\textbf{\ch{N2}} - 1\%}
        \label{fig:N2_2_Kzz6}
    \end{subfigure}
    \hfill
    \begin{subfigure}[b]{0.48\linewidth}
        \includegraphics[width=\linewidth]{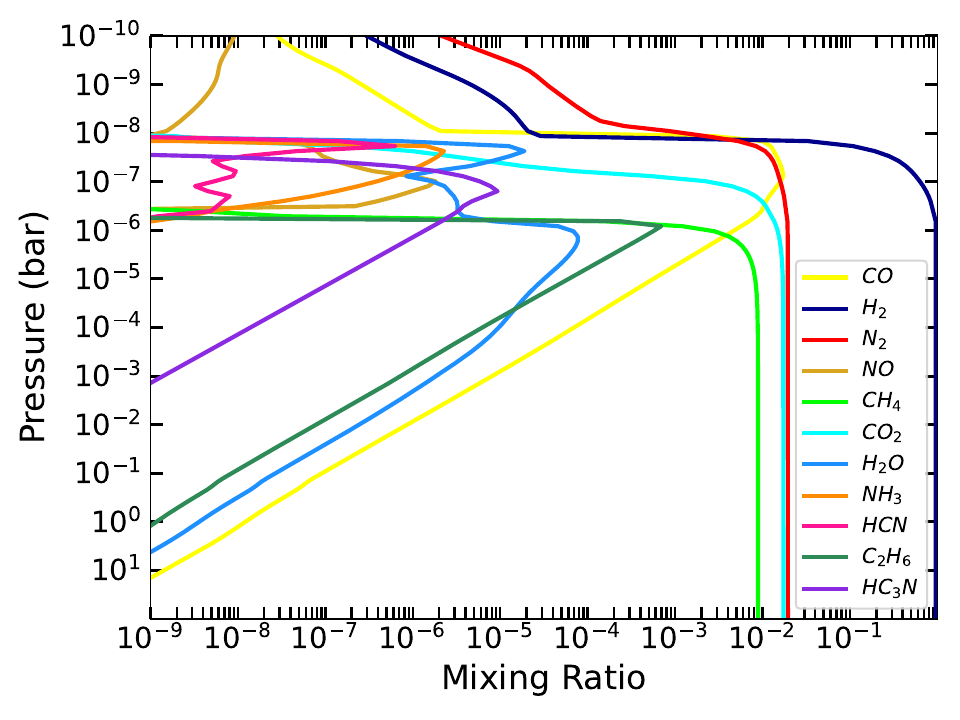}
        \caption{\textbf{\ch{N2}} - 2\%}
        \label{fig:N2_0.02_Kzz6}
    \end{subfigure}

    \vspace{1em} 

    \begin{subfigure}[b]{0.48\linewidth}
        \includegraphics[width=\linewidth]{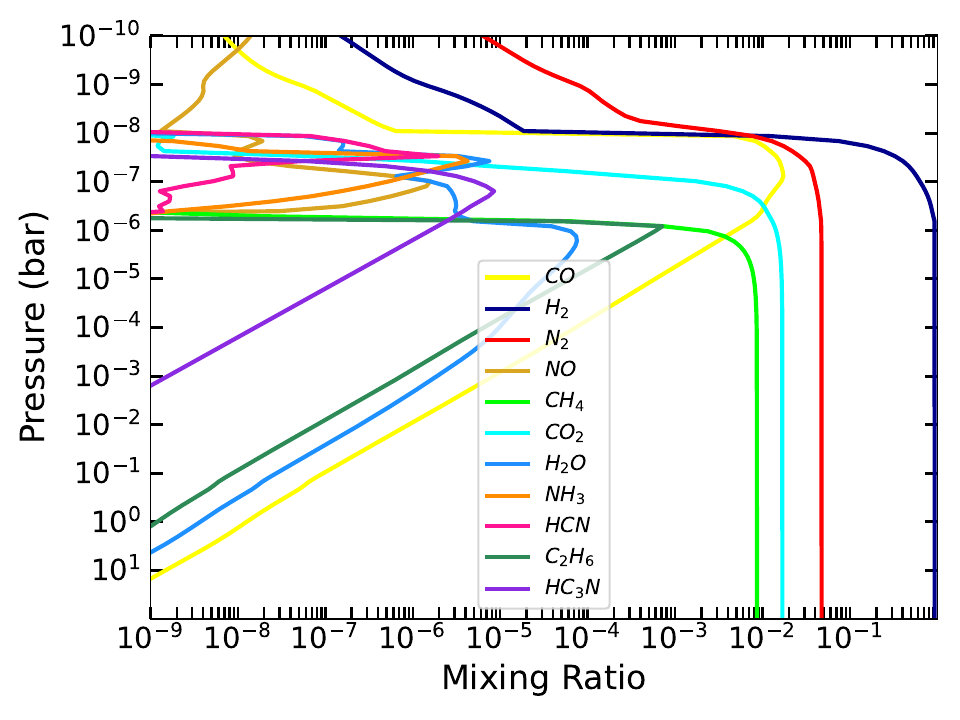}
        \caption{\textbf{\ch{N2}} - 5\%}
        \label{fig:N2_0.05_Kzz6}
    \end{subfigure}%
    \hfill
    \begin{subfigure}[b]{0.48\linewidth}
        \includegraphics[width=\linewidth]{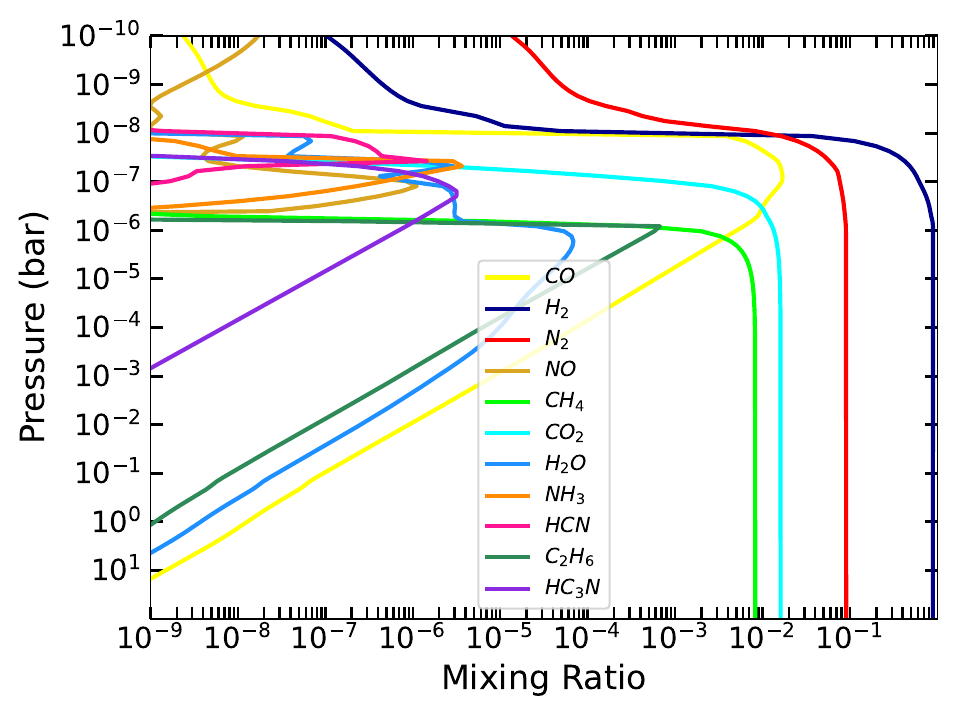}
        \caption{\textbf{\ch{N2}} - 10\%}
        \label{fig:N2_1_Kzz6}
    \end{subfigure}

    \caption{Nitrogen samples with $K_{zz}=10^6$ \unit{cm^2 s^{-1}}. One can see that varying the amount of \ch{N2} has little effect on the overall abundance of nitrogen bearing species. The destruction of \ch{N2} occurs mainly in the upper atmosphere through photolysis - more \ch{N2} initially present at the surface, means more N atoms released at the top of the atmosphere, but not many more nitrogen compounds in the region accessible by our telescopes to view.}
    \label{fig:N2_Kzz_6}
\end{figure*}